\documentclass[showpacs,showkeys,nofootinbib,a4paper,preprint,tightenlines]{revtex4}
\usepackage{amsmath,amssymb,amscd}

\numberwithin{equation}{section}

\newcommand{\al}{\alpha}
\newcommand{\be}{\beta}

\newcommand{\ph}{\varphi}
\newcommand{\ga}{\gamma}
\newcommand{\Ga}{\Gamma}
\newcommand{\de}{\delta}
\newcommand{\la}{\lambda}

\newcommand{\cH}{\mathcal{H}}

\newcommand{\cN}{\mathcal{N}}
\newcommand{\N}{\cN}

\newcommand{\cU}{\mathcal{U}}
\newcommand{\cV}{\mathcal{V}}
\newcommand{\cW}{\mathcal{W}}
\newcommand{\bH}{\mathbf{H}}

\newcommand{\tH}{\tilde{\!H}{}}
 
\newcommand{\tP}{\tilde{P}}
\newcommand{\tcV}{\tilde{\cV}}
\newcommand{\tw}{\tilde w}
\newcommand{\tph}{\tilde{\varphi}}

\newcommand{\Hbp}{\bar{\!H}{}^+}
\newcommand{\bPNp}{\bar P_\N^+}
\newcommand{\bcVp}{\bar{\cV}^+}
\newcommand{\hz}{\hat z}
\newcommand{\hpi}{\hat{\pi}}
\newcommand{\bbN}{\mathbb{N}}
\newcommand{\bbZ}{\mathbb{Z}}
\newcommand{\bbR}{\mathbb{R}}

\newcommand{\fsl}{\mathfrak{sl}}

\newcommand{\ii}{\mathrm i}
\newcommand{\ee}{\mathrm e}
\renewcommand{\d}{\mathrm d}
\newcommand{\Lra}{\Longleftrightarrow}
\newcommand{\sn}{\operatorname{sn}}
\newcommand{\cn}{\operatorname{cn}}
\newcommand{\dn}{\operatorname{dn}}
\newcommand{\pa}{\partial}
\newcommand{\D}{\displaystyle}
\newcommand{\nid}{\noindent}

\begin{document}
\title{A Novel Multi-parameter Family of Quantum Systems\\
  with Partially Broken $\N$-fold Supersymmetry}
\author{Artemio \surname{Gonz\'alez-L\'opez}}
\thanks{email: \texttt{artemio@fis.ucm.es}}
\author{Toshiaki \surname{Tanaka}}
\thanks{email: \texttt{ttanaka@mail.tku.edu.tw}}
\affiliation{Departamento de F\'\i sica Te\'orica II, Facultad de Ciencias F\'\i sicas,
  Universidad Complutense, 28040 Madrid, Spain}
\date{January 31, 2005; revised June 2, 2005}
\begin{abstract}
 We develop a systematic algorithm for constructing an $\N$-fold
 supersymmetric system from a given vector space invariant under one of the
 supercharges. Applying this algorithm to spaces of monomials, we construct a
 new multi-parameter family of $\N$-fold supersymmetric models, which shall be
 referred to as ``type C''. We investigate various aspects of these type C
 models in detail. It turns out that in certain cases these systems exhibit a
novel phenomenon, namely, partial breaking of $\N$-fold supersymmetry. 
\end{abstract}
\maketitle
\section{Introduction}
\label{sec:intro}
A significant progress in scientific research has often been achieved through
the unification of seemingly unrelated concepts. Recently, three different
theoretical developments  were unified in the framework of \emph{$\N$-fold
  supersymmetry} \cite{AST01b,AS03}, namely: i) isospectral transformations,
traced back to the work of Darboux in the late nineteenth century
\cite{Da1882} (see also Ref.~\cite{MS91} and references therein), and their
higher-derivative generalizations first formulated in Ref.~\cite{AIS93}, ii)
quasi-exact solvability in one-dimensional quantum mechanical systems
\cite{TU87} (see also Ref.~\cite{Us94} and references therein), and iii) a
particular class of nonlinear superalgebras. The characteristic feature of
$\N$-fold supersymmetry, which distinguishes it from other nonlinear
extensions of ordinary supersymmetry such as parasupersymmetry
\cite{RS88,BD90,Kh93} and fractional supersymmetry \cite{Du93a}, is the fact
that anticommutators of fermionic operators are polynomials of degree (at
most) $\N$ in bosonic operators. Usually, $\N$-fold supercharges are
represented by $\N$th-order linear differential operators (see references in
Ref.~\cite{Ta03c}). 

The unification mentioned in the previous paragraph comes about schematically
as follows.  An $\N$-fold supersymmetric quantum system always yields, by
definition, a pair of isospectral Hamiltonians $H^\pm$. These operators are
automatically quasi-solvable, since they preserve the kernel of (the bosonic
part of) the respective supercharges $Q^\pm$. Finally, the anticommutator
$\{Q^+,Q^-\}$ is a polynomial of degree (at most) $\N$ in the superHamiltonian
$\bH$, which accounts for the polynomial character of the superalgebra closed
by $\bH$ and $Q^\pm$. Although these ideas are conceptually simple, the direct
construction of $\N$-fold supersymmetric models becomes rather unwieldy for
large $\N$, since intertwining relations with respect to a higher-order
differential operator are quite complicated. Indeed, type A $\N$-fold
supersymmetry \cite{AST01a,Ta03a} is virtually the only known class for which
most of the aspects mentioned above are well understood for arbitrary $\N$. 

One of the characteristic features of type A $\N$-fold supersymmetric models
is that, after a suitable gauge transformation, their supercharges leave
invariant the space of polynomials in one variable of degree less than $\N$. 
In our previous paper \cite{GT03} we obtained a new family of $\cN$-fold
supersymmetric models, the so-called type B, by considering what is probably
the simplest deformation of the type A $\cN$-fold supercharge. Although the
construction presented in this reference was rather \emph{ad hoc} and of a
purely analytic nature, the resulting type B supercharges (after an
appropriate gauge transformation) preserve a finite-dimensional linear space
of monomial type. These results strongly suggest that it can be of great
advantage to base the construction of $\N$-fold supersymmetric models not on a
specific form of the supercharge, but rather on the finite-dimensional linear
space invariant under one of the supercharges, say $Q^-$. 

In this article we show that this idea is indeed feasible, by developing a
systematic algorithm for constructing an $\cN$-fold supersymmetric system
starting from the knowledge of the $\cN$-dim\-ensional linear space of
functions left invariant by a suitable gauge transform of the supercharge
$Q^-$. One of the main advantages of this method is that it completely
bypasses the ``hard'' calculation of the intertwining relations, which makes
it ideally suited when $\N$ is not fixed \emph{a priori}.  We then use our
algorithm to derive the $\cN$-fold supersymmetric systems arising when one
chooses as the starting point in the construction what is perhaps the simplest
possibility, namely a space of monomials.  According to Post and
Turbiner~\cite{PT95}, there exist essentially three inequivalent
finite-dimensional monomial spaces preserved by a nonzero second-order linear
differential operator. It turns out that two of them lead to the already known
$\cN$-fold supersymmetric models of types A and B. The remaining one yields a
new type of supersymmetry, referred to as \emph{type C} in what follows, which
is one of the main contributions of this paper. The type C $\cN$-fold
supercharge is in a sense the most general supercharge of arbitrary order
$\cN$, since it formally reduces to all previously known instances, namely,
the type A and B supercharges. We fully classify the type C models, finding in
particular explicit formulas for their potentials, and further analyze the
main properties of these models, such as shape-invariance, associated
polynomial families of Bender--Dunne type, and the resulting $\cN$-fold
superalgebras. 

The article is organized as follows. In the next section, after reviewing
briefly the concepts of $\cN$-fold supersymmetry and quasi-solvability, as
well as their mutual  relationship, we outline the general procedure for
constructing an $\cN$-fold supersymmetric system starting from a given
finite-dimensional linear space of functions. The connection with the
construction in Refs.~\cite{Ta03a,GT03} of type A and B models is also briefly
discussed.  In Section~\ref{sec:constr} we apply the general method to the
simple case in which the given linear space of functions is of monomial type,
obtaining a new multi-parameter family of $\cN$-fold supersymmetry, namely
type C.  We investigate the general properties of type C models, such as the
shape invariance between the partner Hamiltonians, the structure of the
solvable sectors, and the symmetry transformations which preserve the
potential form.  Using the invariance under these symmetry transformations, we
completely classify the type C models in Section~\ref{sec:class}. We find that
there are essentially four inequivalent nontrivial types of potentials. The
normalizability of the solvable sectors is also briefly examined in this
section.  In Section~\ref{sec:BDpoly}, we study the polynomial families of
Bender--Dunne type associated with the $\cN$-fold supersymmetric models of
type C. We prove that to each type C model one can associate two polynomial
families which, in contrast to their type A counterparts, are always weakly
orthogonal. They also exhibit a novel feature, namely their dependence on two
integer parameters. It is also shown that the polynomial part of the type C
$\cN$-fold superalgebra can be expressed as a product of two critical
polynomials belonging to each of the above families. A few examples are also
exhibited explicitly at the end of this section.  Finally, in Section
\ref{sec:discus}  we discuss several general aspects of type C models, such as
the partial breaking of $\N$-fold supersymmetry, and briefly discuss future
developments suggested by the present work. 

\section{$\N$-fold supersymmetry and quasi-solvability}
\label{sec:NQES}

In this section we shall review the concept of $\N$-fold supersymmetry in
one-dimensional quantum mechanics, making special emphasis on its connections
with the recently introduced notion of quasi-solvability. We shall then
outline a general algorithmic procedure for constructing $\N$-fold
supersymmetric models starting from an $\N$-dimensional linear space invariant
under the action of a second-order linear differential operator. 

Let $q$ denote a bosonic coordinate,
and let $\psi$ and $\psi^\dagger$ be fermionic coordinates satisfying
\begin{align}
  \{\psi,\psi\}=\{\psi^\dagger,\psi^\dagger\}=0, \qquad
  \{\psi,\psi^{\dagger}\}=1. 
\end{align}
Given a monic $\N$th-order linear differential operator
\begin{equation}
\label{eq:Ndfop}
  P_\N =\pa^{\N}_{q}+
    \sum_{k=0}^{\N-1} w_k(q)\,\pa^{k}_{q}\,,
\end{equation}
we introduce the $\N$-fold supercharges $Q_\N ^{\pm}$ by
\begin{align}
\label{eq:dfNsc}
Q_\N ^-=P_\N ^-\psi^{\dagger},\qquad Q_\N ^+=P_\N ^+\psi,
\end{align}
where the operators $P_\N ^\pm$ are defined by
\begin{align}
\label{eq:cmpsc}
P_\N ^-=P_\N ,\qquad P_\N ^+=(-1)^\N P_\N^t. 
\end{align}
The superscript ${}^t$ in the latter equation denotes the transposed
operator defined by $A^t=(A^\dagger)^*$, where the star denotes
complex conjugation. 
If, as is usually the case, all the coefficients $w_k$ in Eq.~\eqref{eq:Ndfop}
are real, then $P_\N^t$ obviously coincides with $P_\N^\dagger$. 
The nilpotency of the fermionic variables $\psi$ and $\psi^\dagger$ implies that
\begin{equation}
\label{eq:QmQp}
\bigl\{Q_\N^-,Q_\N^-\bigr\}=\bigl\{Q_\N^+,Q_\N^+\bigr\}=0\,. 
\end{equation}
We define a \emph{superHamiltonian} $\bH$ by
\begin{align}
\label{eq:NfHam}
\bH=H^-\psi\psi^{\dagger}+H^+\psi^{\dagger}\psi\,,
\end{align}
where
\begin{align}
\label{eq:Ham+-}
H^{\pm}=-\frac12\,\pa_q^2+V^{\pm}(q)
\end{align}
is a pair of scalar Hamiltonians. An \emph{$\N$-fold supersymmetric model}
is a triple $(\bH,Q^+_\N,Q^-_\N)$ such that the supercharges $Q^\pm_\N$ commute
with $\bH$, namely
\begin{equation}
\label{eq:QpmH}
\bigl[Q_\N^\pm,\bH\bigr]=0\,. 
\end{equation}
Equation~\eqref{eq:QpmH} is equivalent to the following \emph{intertwining
 relations} between the component supercharges $P_\N^\pm$ and
Hamiltonians $H^\pm$:
\begin{equation}
  \label{eq:intw}
  P_\N^-H^- - H^+P_\N^-=0, \qquad P_\N^+H^+ - H^-P_\N^+=0\,. 
\end{equation}
Since the Hamiltonians $H^\pm$ are both symmetric under transposition, each
one of the relations \eqref{eq:intw} actually follows from the other one by
transposition. 

Following Ref.~\cite{Ta03a}, we shall say that a differential operator $T$ is
\emph{weakly quasi-solvable} with respect to an $\N$th-order differential
operator $P_\N$ of the form \eqref{eq:Ndfop} if it leaves $\ker P_\N$
invariant.  If, in addition, $\ker P_\N$ can be explicitly computed we shall
simply say that $T$ is quasi-solvable. We shall also use the term
\emph{quasi-exactly solvable} to refer to a quasi-solvable operator whose
finite-dimensional invariant space is a subspace of the Hilbert space on which
the operator is naturally defined.\footnote{Unfortunately, the definition of
  the term ``quasi-exactly solvable'' in the literature is far from uniform. 
  In this paper we have adopted the terminology recently proposed by one of
  the authors in Refs.~\cite{Ta03c,Ta04}.}  A particular class of quasi-solvable
operators is that of \emph{Lie-algebraic} operators, which are polynomials in
the generators of a finite-dimensional Lie algebra of first-order differential
operators preserving a known finite-dimensional linear space
\cite{TU87,Us94,Sh89,GKO94}. 

{}From Eqs.~\eqref{eq:intw} it immediately follows that each one of the
component Hamiltonians $H^\pm$ leaves invariant the kernel $\cV_\N^\pm$ of
the corresponding operator $P_\N^\pm$. Hence both $H^-$ and $H^+$ are
(weakly) quasi-solvable with respect to the operators $P_\N^-$ and $P_\N^+$,
respectively. Remarkably, the converse is also true. Indeed \cite{AST01b},
from a scalar Hamiltonian $H$ weakly quasi-solvable with respect to an
$\N$th-order linear differential operator $P_\N$ of the form \eqref{eq:Ndfop}
one can construct an $\N$-fold supersymmetric model by taking $P_\N^\pm$ as
in \eqref{eq:cmpsc} and
\begin{equation}
  \label{eq:qsNfS}
  H^- = H\,,\qquad H^+=H+\dot w_{\N-1}(q)\,,
\end{equation}
where the dot denotes derivative with respect to $q$. Thus there is
a one-to-one correspondence between $\N$-fold supersymmetric models and
weakly quasi-solvable scalar Hamiltonians. 

{}From the above remarks it follows that, given an $\N$-dimensional linear space
\begin{equation}
\cV_\N=\bigl\langle \ph_1(q),\dots,\ph_\N(q) \bigr\rangle
\label{eq:VN}
\end{equation}
and a scalar Hamiltonian $H$ leaving $\cV_\N$ invariant, one can construct an
$\N$-fold supersymmetric model using Eqs.~\eqref{eq:dfNsc}--\eqref{eq:cmpsc}
and \eqref{eq:qsNfS} and taking as $P_\N$ the unique $\N$th-order linear
differential operator \eqref{eq:Ndfop} annihilating $\cV_\N$, namely
\begin{equation}
P_\N = W(\ph_1,\dots,\ph_\N)^{-1}\,
\begin{vmatrix}
  1& \D\pa_q& \quad\dots\quad & \D\pa_q^\N\\
  \ph_1& \dot\ph_1& \dots & \ph_1^{(\N)}\\
  \vdots& \vdots & \ddots & \vdots\\
  \ph_\N& \dot\ph_\N& \dots & \ph_\N^{(\N)}
\end{vmatrix},
\label{eq:wr}
\end{equation}
where $W(\ph_1,\dots,\ph_\N)$ is the Wronskian of the functions
$\ph_1,\dots,\ph_\N$ spanning $\cV_\N$ (cf.~Refs.~\cite{Cr55,Kr57}). The latter
scheme for constructing $\N$-fold supersymmetric models suffers from two major
drawbacks, namely: i) an arbitrary $\N$-dimensional linear space \eqref{eq:VN}
is not preserved, in general, by any scalar Hamiltonian, and ii) even when
this is the case, it may be difficult to find explicitly a Hamiltonian leaving
\eqref{eq:VN} invariant. 

To overcome these drawbacks, let us slightly generalize the above construction
by considering an $\N$-dimensional linear space
\begin{equation}
\label{eq:VNt}
  \tcV_\N =
  \bigl\langle \tph_1(z),\dots,\tph_\N(z) \bigr\rangle
\end{equation}
and a scalar second-order linear differential operator (not necessarily a
Hamiltonian)
\begin{equation}
  \label{eq:tH}
  -\tH = A(z)\,\pa_z^2+B(z)\,\pa_z+C(z)
\end{equation}
leaving $\tcV_\N$ invariant. {}From these two ingredients one can construct an
$\N$-fold supersymmetric model as follows. 

Let
\begin{equation}
\tP_\N=g(z)\left(\pa^{\N}_{z}+
    \sum_{k=0}^{\N-1} \tw_k(z)\,\pa^{k}_{z}\right)
\label{eq:tPN}
\end{equation}
denote the most general $\N$th-order linear differential operator with
kernel $\tcV_\N$, where the function $g(z)$ is for the time being
undetermined. We shall first construct another second-order linear
differential operator of the form
\begin{equation}
  \label{eq:tHp}
  \tH^+ = \tH - \de C
\end{equation}
satisfying the intertwining relation
\begin{equation}
  \label{eq:itt}
  \tP_\N\,\tH - \tH^+\tP_\N=0\,. 
\end{equation}
To this end, note that the left-hand side of \eqref{eq:itt} is in general a
linear differential operator of order $\N+1$. Equating to zero the
coefficients of $\pa_z^{\N+1}$ and $\pa_z^\N$ in this operator
we obtain the following two equations for the functions $g$ and $\de C$:
\begin{align}
  \frac{g'}g =& \,\frac\N2\,\frac{A'}A
  \label{eq:g}\\
  \de C =& \,\frac12\,\N(\N-2)\left(A''-\frac{{A'}^2}{2A}\right)+
  \N\left(B'-\frac{B\,A'}{2A}\right)\notag\\
  & {}-A'\,\tw_{\N-1}-2A\,\tw_{\N-1}'\,,
  \label{eq:cp}
\end{align}
where the prime denotes derivative with respect to $z$. When
Eqs.~\eqref{eq:g}--\eqref{eq:cp} are satisfied, the l.h.s.\ of \eqref{eq:itt}
is a linear differential operator of order at most $\N-1$ annihilating the
$\N$-dimensional space $\tcV_\N$, and hence it vanishes identically. 

The last step in our construction consists in applying a change of variable
\begin{equation}
  \label{eq:zq}
  z = z(q)
\end{equation}
and a \emph{gauge transformation} with \emph{gauge factor} $\ee^{-\cW(z)}$,
under which
\begin{equation}
  \label{eq:Hpmg}
  \tH^\pm\mapsto
  \left.\ee^{-\cW(z)}\,\tH^\pm\,\ee^{\cW(z)}\right|_{z=z(q)}\equiv H^\pm,
\end{equation}
to simultaneously take $\,\tH^-\equiv\tH$ and $\,\tH^+$ to Schr\"odinger
form \eqref{eq:Ham+-}. Note that this is certainly possible, since (by
construction) $\,\tH$ and $\,\tH^+$ differ by a scalar function only. The
appropriate change of variable and gauge transformation are determined by
\cite{GKO94,GKO93}
\begin{subequations}
  \label{eq:chvgt}
  \begin{align}
    \dot z^2 &= 2\, A(z)
    \label{eq:chv}\\
    \cW' &= \frac1{2A}\Big(\frac{A'}2-B\Big)\,. 
    \label{eq:gt}
  \end{align}
\end{subequations}
The potentials $V^\pm(q)$ are related to the coefficients of the differential
operators $\,\tH^\pm$ as follows:
\begin{equation}
  \label{eq:pots}
  V^\pm(q)=\left.-C^\pm+\frac1{4A}\left[ B^2-A\,A''+\frac34\,{A'}^2
   +2(A B'-A' B)\right]\right|_{z=z(q)},
\end{equation}
where $C^-\equiv C$, $C^+=C^-+\de C$. More explicitly,
\begin{align}
  V^\pm(q) &=
  -C+\frac{A'\tw_{\N-1}}2+A\tw_{\N-1}'
  -\frac12\,(\N-1)\,\left(Q'+\frac{A''}2\right)\notag\\
  &\quad{} +\frac1{4A}\,\left(Q^2+(\N^2-1)\,\frac{{A'}^2}4\right)\notag\\
  &\quad\pm\left.\left[A\tw_{\N-1}'-\frac{\N Q'}2+\Bigl(\frac{\N Q}{4A}
  +\frac{\tw_{\N-1}}2\Bigr)A'\right]\right|_{z=z(q)},
  \label{eq:VpmQ}
\end{align}
where, following Ref.~\cite{GT03}, we have set
\begin{equation}
  \label{eq:Q}
  Q = B+\frac12(\N-2)A'\,. 
\end{equation}
{}From the above construction it immediately follows that the system
\eqref{eq:dfNsc}--\eqref{eq:cmpsc} and \eqref{eq:NfHam},  with $H_\N^\pm$ given
by \eqref{eq:Hpmg} and
\begin{equation}
  \label{eq:PN}
  P_\N=\left.\ee^{-\cW(z)}\,\tP_\N\,\ee^{\cW(z)}\right|_{z=z(q)},
\end{equation}
is $\N$-fold supersymmetric. Indeed, the first intertwining relation
\eqref{eq:intw} follows by applying the gauge transformation and change of
variable \eqref{eq:Hpmg} to the relation \eqref{eq:itt}. Note also that
\eqref{eq:PN} and the definition of $\tP_\N$ imply that the kernel of the
operator $P_\N^-\equiv P_\N$ is given by \eqref{eq:VN}, with
\begin{equation}
  \label{eq:phtph}
  \ph_i(q) = \left.\ee^{-\cW(z)}\,\tph_i(z)\right|_{z=z(q)}\,,
  \qquad 1\leq i\leq\N. 
\end{equation}
Likewise, the invariance of the space $\cV_\N$ under the Hamiltonian $H^-$ is
an immediate consequence of the invariance of $\tcV_\N$ under $\,\tH$ and
Eqs.~\eqref{eq:Hpmg} and \eqref{eq:phtph}. It is important to note that
Eq.~\eqref{eq:PN} for $P_\N$ is compatible with Eq.~\eqref{eq:Ndfop}. Indeed,
from Eqs.~\eqref{eq:g} and \eqref{eq:chv} it follows that $g$ is
proportional to $\dot z^\N$. Taking
\begin{equation}
  \label{eq:gz}
  g(z)=\left.\dot z(q)^\N\right|_{q=q(z)}
\end{equation}
and using Eqs.~\eqref{eq:tPN} and \eqref{eq:PN}, it immediately follows that
$P_\N$ is indeed of the form \eqref{eq:Ndfop}. It is also straightforward to
check, using Eqs.~\eqref{eq:Ndfop}, \eqref{eq:tPN}, \eqref{eq:cp},
\eqref{eq:pots}, \eqref{eq:PN}, and \eqref{eq:gz}, that the partner
Hamiltonians $H^\pm$ are related by \eqref{eq:qsNfS}. 

At this point we shall briefly summarize the results obtained so far. Given an
$\N$-dimensional linear space $\tcV_\N$ \eqref{eq:VNt} and a second-order
linear differential operator $\,\tH\equiv\tH^-$ \eqref{eq:tH} leaving it
invariant, one can construct an $\N$-fold supersymmetric model
$(\bH,Q_\N^\pm)$ through the following algorithmic steps:
\begin{enumerate}
\item Compute the change of variables $z(q)$ using Eq.~\eqref{eq:chv}. 
\item \label{it:step2} Construct the operator
  \begin{equation}
    \label{eq:tPNfinal}
    \tP_N = \dot z^\N \left(\pa^{\N}_{z}+
    \sum_{k=0}^{\N-1} \tw_k(z)\,\pa^{k}_{z}\right)
  \end{equation}
  whose kernel is the linear space $\tcV_\N$. 
\item Compute the operator $\,\tH^+$ \eqref{eq:tHp} using Eq.~\eqref{eq:cp}. 
\item The component Hamiltonians $H^\pm$ are obtained from the \emph{gauged
    Hamiltonians} $\,\tH^\pm$ by applying the gauge transformation
  \eqref{eq:Hpmg}, with $\cW(z)$ given by \eqref{eq:gt}. 
\item Likewise, the operator $P_\N$ determining the $\N$-fold supercharges
  $Q_\N^\pm$ through Eqs.~\eqref{eq:dfNsc}--\eqref{eq:cmpsc} is obtained from
  $\tP_\N$ via the same gauge transformation, cf.~Eq.~\eqref{eq:PN}. 
\end{enumerate}
In practice, the component Hamiltonians $H^\pm$ \eqref{eq:Ham+-} are computed
using Eq.~\eqref{eq:VpmQ} for the potentials $V^\pm(q)$. Note also that the
second step in the previous construction is algorithmic, since the operator in
parenthesis in Eq.~\eqref{eq:tPNfinal} can be computed using a formula
analogous to \eqref{eq:wr}. In most cases, however, the operator $\tP_\N$ is
easily found by inspection. 

To make the above construction completely symmetric with respect to the
partner Hamiltonians $H^-$ and $H^+$, and to facilitate comparison of the
above results with previous work \cite{Ta03a,GT03}, we next introduce
two additional differential operators $\bPNp$ and $\Hbp$ as follows. In the
first place, note that from the second intertwining relation in
Eq.~\eqref{eq:intw} it follows that $H^+$ leaves invariant the kernel of
the supercharge
\begin{equation}
  \label{eq:PNpdef}
  P_\N^+=(-1)^\N P_\N^t=(-1)^\N \ee^\cW\tP_\N^t\,\ee^{-\cW}. 
\end{equation}
Using the identity\footnote{Note that the transposition has been defined in
terms of the variable $q$, cf.~Eq.~\eqref{eq:cmpsc}.} 
$$
\bigl(\pa_z\bigr)^t = \left(\frac1{\dot z}\,\pa_q\right)^t
=-\pa_q\,\frac1{\dot z}=-\dot z \,\pa_z\,\frac1{\dot z}
$$
and Eq.~\eqref{eq:tPNfinal} we can write
\begin{equation}
  \label{eq:tPNad}
  (-1)^\N \tP_N^t = \dot z^{1-\N}\bPNp\,\dot z^{\N-1}\,,
\end{equation}
where
\begin{equation}
  \label{eq:bPNp}
  \bPNp = \dot z^\N\left(\pa_z^\N+\sum_{k=0}^{\N-1}(-1)^{\N-k}
  \pa_z^k\,\tw_k\right)\,. 
\end{equation}
{}From Eqs.~\eqref{eq:PNpdef} and \eqref{eq:tPNad} we have
\begin{equation}
  P_\N^+=\ee^{-\cW^+}\,\bPNp\,\ee^{\cW^+}\,,
  \label{eq:PbP}
\end{equation}
where the function $\cW^+$ is given by
\begin{equation}
  \label{eq:cWp}
  \cW^+ = -\cW+(\N-1)\ln|\dot z|\,. 
\end{equation}
The invariance of $\ker P_N^+$ under $H^+$ and Eq.~\eqref{eq:PbP} imply that
the operator
\begin{equation}
  \label{eq:Hbp}
  \Hbp = \ee^{\cW^+}\,H^+\ee^{-\cW^+}
\end{equation}
leaves the linear space
\begin{equation}
  \label{eq:VNbar}
  \bar\cV_\N^+ = \ker\bPNp
\end{equation}
invariant. Setting $\cW^-\equiv\cW$ we can express the partner Hamiltonians as
\begin{equation}
H^\pm=\ee^{-\cW^\pm}\,\bar{\tilde H}^\pm \ee^{\cW^\pm}\,,
\label{eq:Hpmgauge}
\end{equation}
where the gauged Hamiltonians $\bar{\tilde H}^\pm$ leave invariant the
kernel of the ``gauged'' supercharges $\bar{\tilde P}_\N^\pm$ respectively
given by Eqs.~\eqref{eq:tPNfinal} and \eqref{eq:bPNp}. The ``physical''
supercharges $P^\pm_\N$ are related to the gauged supercharges by the
equations
\begin{equation}
  \label{eq:Ps}
  P^\pm_\N=\ee^{-\cW^\pm}\,\bar{\tilde P}_\N^\pm \ee^{\cW^\pm}\,. 
\end{equation}
In order to express $\cW^\pm$ in a symmetric way, we introduce
the functions
\begin{equation}
  \label{eq:Wq}
  W(q)=\frac12\left(\dot{\cW}^- (q)-\dot{\cW}^+ (q)\right)
\end{equation}
and
\begin{equation}
  \label{eq:E}
  E(q)=\frac{\ddot{z}(q)}{\dot{z}(q)}\,. 
\end{equation}
{}From Eq.~\eqref{eq:chv}, its immediate consequence
\begin{equation*}
  \ddot{z}=A'\,,
\end{equation*}
and Eq.~\eqref{eq:gt} it is straightforward to derive the relation
\begin{equation}
  \label{eq:W}
  W=-\frac{Q}{\dot{z}}\,. 
\end{equation}
We then have
\begin{equation}
  \label{eq:cWpm}
  \cW^\pm =\frac12(\N-1)\int E(q)\,\d q\mp\int W(q)\,\d q
  =\frac14(\N-1)\ln\bigl|2A(z)\bigr|\pm\int\frac{Q(z)}{2A(z)}\,\d z\,,
\end{equation}
cf.~\cite{GT03}. The connection between the gauged Hamiltonians
$\Hbp$ and $\,\tH^+$ follows
easily from Eqs.~\eqref{eq:Hpmg} and \eqref{eq:Hbp}, namely
\begin{equation}
  \label{eq:Hbt}
  \Hbp = \ee^{-2\int W(q)\d q}\,\tH^+\,\ee^{2\int W(q)\d q}\,. 
\end{equation}
Using Eqs.~\eqref{eq:cp}, \eqref{eq:chv}, \eqref{eq:Q}, and \eqref{eq:W}, it
is immediate to obtain the following explicit expression for $\Hbp$:
\begin{equation}
  \label{eq:Hbpexp}
  -\Hbp=A\,\pa_z^2-\left(Q+\frac12(\N-2)A'\right)\,\pa_z
  +C+(\N-1)Q'-A'\tw_{\N-1}-2A\,\tw'_{\N-1}\,. 
\end{equation}
Combining this equation with Eq.~\eqref{eq:tH} we obtain the following
unified formula for the gauged Hamiltonians $\bar{\tilde H}^\pm$:
 \begin{align}
   -\bar{\tilde H}^\pm=A\,\pa_z^2&-\left(\pm
   Q+\frac12(\N-2)A'\right)\,\pa_z+C\notag\\
   &+\frac12(1\pm1)\left((\N-1)Q'-A'\tw_{\N-1}
   -2A\,\tw'_{\N-1}\right)\,. 
   \label{eq:Hgpmcoll}
 \end{align}
This general formula includes as particular cases the gauged
Hamiltonians of the type A and type B models introduced respectively in
Refs.~\cite{Ta03a} and \cite{GT03}.  Indeed, in the type A models we have
$\tw_{\N-1}=0$ and
$$
C = \frac1{12}(\N-1)(\N-2)A''-\frac12(\N-1)Q'+R\,,
$$
where $R$ is a constant. Using these relations in Eq.~\eqref{eq:Hgpmcoll}
we can easily reproduce Eqs.~(3.41) and (3.50b), (3.52), (3.55) of
Ref.~\cite{Ta03a} for the gauged Hamiltonians of type A (note that the
coordinate $h$ and the function $P$ in the latter reference correspond to our
$z$ and $A$, respectively). Likewise, type B models satisfy $\tw_{\N-1}=-1/z$
and
$$
C = \frac1{12}(\N-1)(\N-2)A''-\frac{A'}{2z}-\frac12(\N-1)Q'-\frac Q{\N z}+R\,,
$$
which immediately lead to Eq.~(3.35) of Ref.~\cite{GT03} for the gauged
Hamiltonians of type B. 

\section{A new family of $\N$-fold supersymmetric systems}
\label{sec:constr}

In this section we shall apply the previous results to the construction of a
new multi-parameter family of $\N$-fold supersymmetric systems for arbitrary
$\N$. The key idea in this respect is to choose appropriately
the $\N$-dimensional linear space \eqref{eq:VNt}, in such a way that the linear
space of second-order linear differential operators leaving it invariant is
nontrivial and can be explicitly computed. 

To this end, we shall consider \emph{monomial spaces} of the form
\begin{equation}
  \label{eq:mon}
  \tcV_\N=\bigl\langle z^{\la_1},\dots,z^{\la_\N}\bigr\rangle,
\end{equation}
where the exponents $\la_i$ are real numbers. All spaces \eqref{eq:mon}  left
invariant by a nonzero second-order linear differential operator have been
classified by Post and Turbiner \cite{PT95}, up to changes of variables and
gauge transformations of the form
\begin{equation}
\psi(z)\mapsto\hat\psi(\hz)=\hz^\al\,\psi(\hz^\be)\,,\qquad
\al,\be\in\bbR\,.\label{eq:pgt}
\end{equation}
For $\N>4$ the above classification consists of three equivalence classes,
represented by the following ``canonical forms'':
\begin{align}
  \label{eq:typeA}
  \text{A)}\quad&\bigl\langle 1,z,\dots,z^{\N-1}\bigr\rangle\\
  \label{eq:typeB}
  \text{B)}\quad&\bigl\langle 1,z,\dots,z^{\N-2},z^\N\bigr\rangle\\
  \label{eq:typeC}
  \text{C)}\quad&\bigl\langle 1,z,\dots,z^{\N_1-1},z^\la,z^{\la+1},\dots,
  z^{\la+\N_2-1}\bigr\rangle,\qquad \N=\N_1+\N_2\,. 
\end{align}
In the third canonical form $\N_1$ and $\N_2$ are positive integers, and $\la$
is a real number different from $-\N_2,-\N_2+1,\dots,\N_1$
($\la\neq-\N_2-1,\N_1+1$ when $\N_1=1$ or $\N_2=1$ to prevent the canonical form
C from reducing to B).  Due to the freedom in performing changes of
variables and gauge transformations \eqref{eq:pgt} we can also assume, without
loss of generality, that
\begin{equation}
  \label{eq:rest}
  \la>0\,,\qquad \N_1\geq\N_2\,. 
\end{equation}

The $\N$-fold supersymmetric models constructed from the canonical form A
following the procedure described in the previous section are nothing but the
type A systems introduced in Ref.~\cite{AST01a}. Similarly, the
models obtained from the second canonical form are the type B systems
recently constructed by the authors~\cite{GT03}. In this section we shall
therefore derive and completely classify all the $\N$-fold supersymmetric
models associated to the last canonical form \eqref{eq:typeC}, which from
now on we shall term \emph{type C} models for short. 

We shall now proceed to the construction of the type C models by following the
algorithmic steps outlined in Section~\ref{sec:NQES}. In order to implement
this algorithm, we must know at least one nonzero second-order linear
differential operator leaving the space~\eqref{eq:typeC} invariant. Due to the
extremely simple nature of the latter space, it is actually straightforward to
determine the whole linear space of linear differential operators of
order not greater than two preserving it. Indeed, using the techniques
described in Ref.~\cite{PT95} it is readily found that the latter space is
spanned by the constant multiplication operator $1$ and the following
operators:
\begin{subequations}
  \label{eq:Js}
  \begin{align}
  \label{eq:J0m}
  J_{0-} &= \pa_z\bigl(z\pa_z-\la\bigr)\\
  \label{eq:J00}
  J_{00} &= z^2\,\pa_z^2\\
  \label{eq:Jp0}
  J_{+0} &= z\bigl(z\pa_z-\N_1+1\bigr)
  \bigl(z\pa_z-\la-\N_2+1\bigr)\\
  \label{eq:J0}
  J_0 &= z\,\pa_z\,. 
  \end{align}
\end{subequations}
The most general linear second-order differential operator leaving the space
\eqref{eq:typeC} invariant is thus given by
\begin{equation}
  \label{eq:Hg}
  -\tH = a_1 J_{0-}+a_2 J_{00}+a_3J_{+0}+b_0 J_0+c_0\,,
\end{equation}
where the coefficients $a_i$, $b_0$, and $c_0$ are real constants.  More
explicitly, the coefficients $A(z)$, $B(z)$ and $C(z)$ of $-\tH$
(cf.~\eqref{eq:tH}) are given by
\begin{subequations}
  \label{eq:abcz}
  \begin{align}
    \label{eq:az}
    A(z) &= a_3 z^3 + a_2 z^2+ a_1 z\\
    \label{eq:bz}
    B(z) &= -
    (\N +\la-3)\,a_3 z^2 + b_0 z + (1-\la)\,a_1\\
    \label{eq:cz}
    C(z) &= (\N_1-1)(\N_2+\la-1)\,a_3z+c_0\,. 
  \end{align}
\end{subequations}
By Eq.~\eqref{eq:Q} we have
\begin{equation}
  \label{eq:Q-C}
  Q=\frac12(\N-2\la)\,a_3 z^2+b_1\,z+\frac12(\N-2\la)\,a_1,
\end{equation}
with
\begin{equation}
  \label{eq:b1}
  b_1 = b_0 + (\N-2)\,a_2\,. 
\end{equation}
{}From Eqs.~\eqref{eq:chv} and \eqref{eq:az}, the change of variable $z(q)$
is determined in this case by the differential equation
\begin{equation}
  \label{eq:zqde}
  \dot z^2 = 2\bigl(a_3 z^3 + a_2 z^2 + a_1 z\bigr). 
\end{equation}

The type C space \eqref{eq:typeC} decomposes in a natural way as the direct
sum of two spaces of type A, namely
\begin{equation}
  \label{eq:csum}
  \bigl\langle 1,z,\dots,z^{\N_1-1}\bigr\rangle\oplus z^\la\bigl\langle 1,z,
  \dots,z^{\N_2-1}\bigr\rangle\,. 
\end{equation}
It is important to observe that both subspaces in the latter sum are
\emph{separately} invariant under all of the operators \eqref{eq:Js}, and
hence under the gauged Hamiltonian $\,\tH$. In particular, a result of
Turbiner \cite{Tu94} implies that $\,\tH$ and $z^\la\,\tH\,z^{-\la}$ are
Lie-algebraic operators with respect to the standard realization of the
algebra $\fsl(2)$ with generators
\begin{equation}
  \label{eq:sl2}
  J_-=\pa_z\,,\qquad J_0=z\,\pa_z\,,\qquad J_+=z^2\pa_z-n\,z
\end{equation}
and cohomology parameter $n=\N_1-1$ and $n=\N_2-1$, respectively (this
property can also be checked directly using \eqref{eq:Js}).  However, an
arbitrary polynomial in the operators \eqref{eq:sl2} will not preserve, in
general, \emph{both} type A spaces in \eqref{eq:csum}. For this reason, the
number of independent first- and second-order operators preserving $\tcV_\N$
is reduced from 8 for type A to 4 for type C. In particular, the set of type C
gauged Hamiltonians does not include all gauged Hamiltonians of type A. In
Section~\ref{sec:BDpoly} we will discuss this phenomenon from a different
viewpoint, namely the breakdown of an underlying symmetry.  It should also be
noted in this respect that the gauged Hamiltonian of a type B model is not, in
general, a polynomial in the $\fsl(2)$ generators~\eqref{eq:sl2}. 

The operators $J_{0-}$, $J_{00}$, and $J_0$ obviously
leave invariant the spaces $\langle 1,z,\dots,z^{\N_1-1}\rangle$
and $z^\la\langle 1,z,\dots,z^{\N_2-1}\rangle$ for
\emph{arbitrary} positive-integer values of $\N_1$ and $\N_2$. 
It follows from Eq.~\eqref{eq:Hg} that the gauged Hamiltonian $\,\tH$
will also leave invariant the latter spaces for all $\N_1,\N_2\in\bbN$
provided that the coefficient $a_3$ vanishes. When this is the case,
Eq.~\eqref{eq:Hpmgauge} implies that the Hamiltonian $H^-$ preserves
the two infinite ascending sequences of spaces
\[
\cV_{\N_i}^-\equiv\ee^{-\cW^-(q)}z(q)^{(i-1)\la}\bigl\langle1,z(q),
 \dots,z(q)^{\N_i-1}\bigr\rangle\,;\qquad i=1,2\,,\quad\N_i\in\bbN\,. 
\]
Hence for $a_3=0$ the type C component Hamiltonian $H^-$ is \emph{solvable} in
Turbiner's sense \cite{Tu88,Tu92} (see also \cite{Ta03c,Ta04}). We will see
shortly that in this case the other Hamiltonian $H^+$ is simultaneously
solvable, as for type A models. 

The $\N$th-order linear differential operator $\tP_\N$ of the form
\eqref{eq:tPNfinal} having as kernel the type C space \eqref{eq:typeC} is
easily found to be
\begin{equation}
  \tP_\N = \dot z^\N
  \left(\pa_z+\frac{\N_1-\la}z\right)^{\N_2}\pa_z^{\N_1}. 
  \label{eq:tPNC}
\end{equation}
In particular, the function $\tw_{\N-1}$ is given in this case by
\begin{equation}
  \label{eq:twgen}
  \tw_{\N-1}=\frac{\N_2(\N_1-\la)}z\,. 
\end{equation}

To compute the operator $P_\N$ determining the supercharges $Q_\N^\pm$, let
us introduce the function
\begin{equation}
  \label{eq:F}
  F(q) = \frac{\dot z(q)}{z(q)},
\end{equation}
related to $E$ by the identity (see Eq.~\eqref{eq:E})
\begin{equation}
\label{eq:EFid}
\dot F = E\,F-F^2\,. 
\end{equation}
Making repeated use of the equality
\begin{equation}
  \label{eq:idPN}
  \left(\pa_z+\frac\al z\right)\dot z^{-k} =
  \dot z^{-1}\bigl(\pa_q+\al F\bigr)\dot z^{-k}
  = \dot z^{-k-1}\bigl(\pa_q+\al F-k E\bigr)
\end{equation}
we immediately obtain
\begin{equation}
  \label{eq:tPNq}
  \tP_\N = \prod_{i=\N_1}^{\N-1}\bigl(\pa_q+(\N_1-\la)F-i E\bigr)
  \cdot \prod_{i=0}^{\N_1-1}\bigl(\pa_q-i E\bigr)\,,
\end{equation}
where the products of operators are ordered according to the following
definition:
$$
\prod_{i=i_0}^{i_1} A_i \equiv A_{i_1}A_{i_1-1}\cdots A_{i_0}\,. 
$$
Using the identity
\begin{equation}
  \ee^{-\cW}\pa_q\,\ee^{\cW} = \pa_q+\dot\cW = 
  \pa_q+\frac12(\N-1) E(q)+W(q)
  \label{eq:gaugeW}
\end{equation}
and Eq.~\eqref{eq:PN} we finally obtain the following explicit formula for the
operator $P_\N$ of type C:
\begin{multline}
  \label{eq:PNC}
  P_\N = \prod_{i=\N_1}^{\N-1}\left(\pa_q+W+(\N_1-\la)F
  +\frac12\,(\N-1-2i)E\right)\\
  \times \prod_{i=0}^{\N_1-1}\left(\pa_q+W+\frac12\,(\N-1-2i)E\right)\,. 
\end{multline}
It is clear from the previous expression that the supercharge \eqref{eq:PNC}
reduces to its type A counterpart for $\la=\N_1=\N-1$, and to the type B one
for $\la=\N_1+1=\N$. We can thus formally regard the type C supercharge as a
deformation of those of types A and B depending on the two parameters
$\N_1$ and $\la$. 

The pair of type C potentials $V^\pm$ can be expressed in terms of
the functions $E$, $F$ and $W$ using the following identities, which
are easily derived from Eqs.~\eqref{eq:chv}, \eqref{eq:E}, \eqref{eq:W},
\eqref{eq:twgen}, and \eqref{eq:F}:
\begin{gather}
A'\tw_{\cN-1}=\cN_2(\cN_1-\la)EF\,,\qquad
 2A\tw_{\cN-1}'=-\cN_2(\cN_1-\la)F^2,\\
\begin{aligned}
2A&=z^2 F^2, \quad & \frac{A'^2}{2A}&=E^2, &  A''&=\dot E+E^2,\\
\frac{Q^2}{2A}&=W^2, & \frac{QA'}{2A}&=-EW, \quad &
Q'&=-(\dot W+E\,W)\,. 
\end{aligned}
\end{gather}
Substituting the above formulas into Eq.~\eqref{eq:VpmQ} and
using the relation \eqref{eq:EFid}, we finally obtain\footnote{Due to
the relation \eqref{eq:EFid}, this expression is not unique.} 
\[
 H^{\pm}=-\frac12\,\pa_q^2+\frac{W^2}2
  -\frac{\cN_2(\cN_1-\la)}4\,F^2-\frac{\cN^2-1}{24}
  \left(2\dot E-E^2\right)
 \pm\left[\frac{\cN}2\,\dot W +\frac{\cN_2(\cN_1-\la)}2\,\dot F\right]
  -R\,,
\]
where $R$ is a constant. 

{}From Eqs.~\eqref{eq:bPNp} and \eqref{eq:tPNC} it follows that the operator
$\bPNp$ is given in this case by
\begin{equation}
  \label{eq:bPNpC}
  \bPNp=\dot z^\N
  \pa_z^{\N_1}\left(\pa_z-
  \frac{\N_1-\la}z\right)^{\N_2}. 
\end{equation}
Its kernel $\bcVp_\N$ is easily computed (cf.~Eq.~\eqref{eq:VNbar}), with the
result
\begin{align}
  \label{eq:bPNpcan}
  \bcVp_\N=z^{\N_2}\bigl\langle1,z,\dots,z^{
    \N_1-1},z^{\bar\la},z^{\bar\la+1},\dots,z^{\bar\la+\N_2-1}\bigr\rangle,
\end{align}
where
\begin{equation}
  \label{eq:bla}
  \bar\la=\N_1-\N_2-\la\,. 
\end{equation}
The space $\bcVp_\N$ can obviously be transformed into the type C canonical
form \eqref{eq:typeC} by the gauge transformation $\psi(z)\mapsto \hat\psi(z)=
z^{-\N_2}\psi(z)$.  Since $\bar\la$ is negative if $\la>\N_1-\N_2$, strictly
speaking it is necessary in this case to perform an additional gauge
transformation and change of variable \eqref{eq:pgt} to take the space
\eqref{eq:bPNpcan} into the canonical form
\eqref{eq:typeC}--\eqref{eq:rest}. It is preferable, however, not to enforce
the first restriction $\la>0$ in Eq.~\eqref{eq:rest} in the sequel (cf.~the
discussion following Eq.~\eqref{eq:proj}), the positivity of $\la$ being
immaterial in other respects. Hence from now on we shall only impose on $\la$
the restriction needed to ensure that the type C space \eqref{eq:typeC} does
not reduce to the type A or B canonical forms, namely
\begin{equation}
  \label{eq:restla}
  \la\in\bbR\setminus\{-\N_2,-\N_2+1,\dots,\N_1\}\,,
\end{equation}
with $\la\neq-\N_2-1,\N_1+1$ if $\N_1=1$ or $\N_2=1$. It is easily seen from
 Eq.~\eqref{eq:bla} that $\bar{\la}$ is restricted in exactly the same way as $\la$. 

{}From Eq.~\eqref{eq:bPNpcan} it follows that the gauged Hamiltonians
$\Hbp[a_i,b_1,c_0,\la]$ and $\,\tH^-[\bar a_i,\bar b_1,\bar c_0,\bar\la]$
are related by the gauge transformation
 \begin{equation}
   \label{eq:relgHpm}
   \Hbp[a_i,b_1,c_0,\la]=
   z^{\N_2}\tilde H^-[\bar a_i,\bar b_1,\bar c_0,\bar\la]\,z^{-\N_2}
 \end{equation}
for a suitable choice of the parameters $\bar a_i$, $\bar b_1$, and $\bar
c_0$. The latter equality implies, by the uniqueness (up to an additive
constant) of the physical Hamiltonian associated to a given gauged
Hamiltonian, the important relation
\begin{equation}
  \label{eq:relHpm}
  H^+[a_i,b_1,c_0,\la]=H^-[\bar a_i,\bar b_1,\bar c_0,\bar\la]\,. 
\end{equation}
In other words, the type C $\N$-fold supersymmetric models we shall obtain
are guaranteed to be (formally) \emph{shape invariant} \cite{CKS95}. 
Explicitly, the parameters $(\bar{a}_i,\bar{b}_1,\bar{c}_0)$ are
given by,
\begin{subequations}
\begin{align}
  \bar{a}_i &= a_i,\qquad i=1,2,3,\\
  \bar{b}_1 &= -b_1+2\N_2 a_2,\\
  \bar{c}_0 &= c_0 +(\N_1-1)(b_1 -\N_2 a_2). 
\end{align}
\end{subequations}
Since $a_3=0$ implies $\bar{a}_3=0$ and vice versa, it follows that
$H^-$ and $H^+$ are always simultaneously solvable. 

We shall next examine the explicit forms of the
subspaces $\cV^{\pm}_{\N}$ preserved by the type C Hamiltonians
$H^{\pm}$. We first note from Eqs.~\eqref{eq:az} and \eqref{eq:Q-C}
that the second term of the last expression for $\cW^\pm$ in
Eq.~\eqref{eq:cWpm} is given by
\begin{equation}
  \int\frac{Q(z)}{2A(z)}\,\d z=\frac{\N-2\la}4\ln |z|
  +\left(\frac{b_1}2-\frac{\N-2\la}4a_2\right)\int\frac{z}{A(z)}\,\d z\,. 
\end{equation}
Introducing the new parameters
\begin{equation}
  \label{eq:lapm}
  \al^- =  \frac12-\la\,,\qquad \al^+ = \N_2 -\N_1+\la +\frac12
  =\frac12-\bar{\la}\,,
\end{equation}
we obtain the following expression for the gauge factors of the type C models:
\begin{multline}
  \label{eq:cWpm-C}
  \cW^\pm =-\left(\frac{\al^\pm}2-\frac{1\pm 1}2 \N_2\right)\ln |z|
  +\frac{\N-1}4 \ln \left|\frac{A(z)}z\right|\\
  \pm\left(\frac{b_1}2-\frac{\N-2\la}4 a_2\right)
  \int\frac{z}{A(z)}\,\d z\,,
\end{multline}
where an irrelevant constant term has been dropped.  By
Eq.~\eqref{eq:Hpmgauge}, the subspaces $\cV^\pm_\N$ invariant under the
Hamiltonians $H^\pm$, which provide the algebraically computable wave functions
(not taking their normalizability into account), are obviously given by
\begin{equation}
  \label{eq:modules}
  \cV^\pm_\N=\ee^{-\cW_\pm}\,\bar{\tcV}^\pm_\N\,. 
\end{equation}
{}From Eqs.~\eqref{eq:typeC},
\eqref{eq:bPNpcan} and \eqref{eq:cWpm-C} we finally have
\begin{align}
  \label{eq:cVNpm}
  \cV_\N^\pm &= \left.\ee^{-\cU^\pm}z^{\frac12\al^\pm}\left\langle 1,z,\dots,
   z^{\N_1 -1}\right\rangle\oplus\ee^{-\cU^\pm}z^{\frac12(1-\al^\pm)}
   \left\langle 1,z,\dots,z^{\N_2 -1}\right\rangle\right|_{z=z(q)}\notag\\
  &\equiv\cV_{\N_1}^\pm \oplus\cV_{\N_2}^\pm,
\end{align}
where the new gauge factors $\ee^{-\cU^\pm}$ are defined by
\begin{equation}
  \ee^{-\cU^\pm}=\left|\frac{A(z)}z\right|^{-\frac14(\N-1)}
  \exp\left[\mp\left(\frac{b_1}2-\frac{\N-2\la}4 a_2\right)
  \int\frac{z}{A(z)}\,\d z\right]\,. 
\end{equation}

The last step in our construction is the computation of the component
Hamiltonians $H^\pm$ using Eqs.~\eqref{eq:Ham+-} and \eqref{eq:VpmQ}. This is,
technically speaking, the most delicate step, since it involves the explicit
evaluation of the elliptic integral
\begin{equation}
  \label{eq:ellint}
  \int\bigl[2(a_3 z^3+a_2z^2+a_1z)\bigr]^{-1/2}\d z = \pm(q -q_0)
\end{equation}
needed to compute the change of variable $z=z(q)$, cf.~Eq.~\eqref{eq:zqde}. 

The value of the integral \eqref{eq:ellint}, and hence the corresponding
change of variable, depends on the position of the roots of the polynomial
$A(z)$ in the complex plane. We should therefore classify the polynomial
$A(z)$ into (real) canonical forms according to the position of its roots,
using changes of variables and gauge transformations
\begin{equation}
  \label{eq:classtr}
  \psi(z)\mapsto\hat\psi(\hz)=\mu(z)\psi(z)\bigr|_{z=\zeta(\hz)}
\end{equation}
that preserve the form of the type C space~\eqref{eq:typeC}. This task is
hindered by the fact that, contrary to what happens in the analogous
classification of the one-dimensional Lie-algebraic and type A $\N$-fold
supersymmetric Hamiltonians~\cite{Ta03a,GKO94,GKO93}, the
space~\eqref{eq:typeC} is not invariant under translations $z=\hz+z_0$. 
Fortunately, however, this space is invariant under dilations
\begin{equation}
  \label{eq:dil}
  \psi(z)\mapsto\hat\psi(\hz) \equiv \psi(\al \hz)\,,\qquad\al\in\bbR\,,
\end{equation}
and is form-invariant under \emph{special projective transformations}
\begin{equation}
  \label{eq:proj}
  \psi(z)\mapsto\hat\psi(\hz) \equiv \hz^s\,\psi(\hz^{-1})\,,
  \qquad s=\N_1-1,\;\N_2+\la-1\,. 
\end{equation}
Since, however, neither projective transformation \eqref{eq:proj} preserves
both conditions \eqref{eq:rest}, we must drop one of these conditions if we
insist on using projective transformations to bring $A(z)$ into canonical
form. {}From now on we shall assume that only the second condition $\N_1\geq N_2$ in
Eq.~\eqref{eq:rest} holds,
which entails the choice $s=\N_1-1$ in Eq.~\eqref{eq:proj}. 

The transformations \eqref{eq:dil}--\eqref{eq:proj} map the gauged Hamiltonian
$\,\tH$ \eqref{eq:tH} into the operator $\hat H$ respectively given by
$$
-\hat H = \left.-\tH\right|_{z=\al \hz} = \frac1{\al^2}\,A(\al
\hz)\,\frac{\d^2}{\d \hz^2} +\frac1\al\,B(\al \hz)\,\frac\d{\d \hz} + C(\al
\hz)
$$
and
\[
  -\hat H = \left.-\hz^s\,\tH\,\hz^{-s}\right|_{z=\hz^{-1}}
  = A(\hz^{-1})\,\Big(-\hz^2\frac\d{\d \hz}+s\,\hz\Big)^{\!2} +
  B(\hz^{-1})\,\Big(-\hz^2\frac\d{\d \hz}+s\,\hz\Big)+C(\hz^{-1})\,. 
\]
In particular, the polynomial $A(z)$ transforms under dilations and special
projective transformations respectively as
\begin{equation}
  \label{eq:adil}
  A(z)\mapsto\hat A(\hz)\equiv\frac1{\al^2}\,A(\al \hz)
\end{equation}
and
\begin{equation}
  \label{eq:apr}
  A(z)\mapsto\hat A(\hz)\equiv \hz^4 A(\hz^{-1})\,. 
\end{equation}
With the help of the transformations \eqref{eq:adil}--\eqref{eq:apr} it is
readily shown that $A(z)$ can be cast into one of the canonical forms listed
in Table~\ref{tb:table1}. 
\begin{table}
  \caption{Canonical forms for the polynomial $A(z)$ \eqref{eq:az}. In this Table $\nu>0$, $0\leq m\leq 1$ and $m'=1-m$ ($m\neq1$ in Case 3 and
    $m\neq0,1$ in Case 4 to avoid duplications).
  } 
  \label{tb:table1}
  \vspace*{.6cm}
  \begin{center}
    \begin{tabular}{ll}
      \hline
      \hline
      1)\qquad & $2z$\\
      2)\qquad & $\pm\frac12\,\nu z^2$\\
      3)\qquad & $\pm2\nu z(1-z)(1-m z)$\\ 
      4)\qquad & $2\nu z(1-z)(m'+m z)$\\
      5)\qquad & $\frac12\,\nu z\bigl(z^2+2(1-2m)z+1\bigr)$\\[1mm]
      \hline
      \hline
    \end{tabular}
  \end{center}
  \vspace*{.6cm}
\end{table}
The discussion following Eq.~\eqref{eq:zqde} implies that 
the type C models corresponding to the first two canonical forms
in Table \ref{tb:table1}, or to the third one with $m=0$,
are not only quasi-solvable but also solvable. 

\section{Classification of the type C models}
\label{sec:class}
We shall now explicitly compute the type C models associated to each of the
canonical forms in Table \ref{tb:table1}. Note that, by Eq.~\eqref{eq:zqde},
a rescaling of the coefficients $a_i$, $b_1$, $c_0$ by an overall nonzero constant
factor $\nu$ has the following effect on the change of variable $z(q)$:
\begin{equation}
  \label{eq:zscale}
  z(q;\nu a_i,\nu b_1,\nu c_0)=z(\sqrt\nu\,q;a_i,b_1,c_0)\,. 
\end{equation}
{}From this equation and Eqs.~\eqref{eq:VpmQ}, \eqref{eq:Q},
\eqref{eq:E}--\eqref{eq:cWpm}, and \eqref{eq:F} we easily
obtain the identities
\begin{subequations}
\label{eq:EFWV}
\begin{align}
  \label{eq:Escale}
  E(q;\nu a_i,\nu b_1,\nu c_0) &=
  \sqrt\nu\,E(\sqrt\nu\,q;a_i,b_1,c_0)\\[1pt]
  \label{eq:Fscale}
  F(q;\nu a_i,\nu b_1,\nu c_0) &=
  \sqrt\nu\,F(\sqrt\nu\,q;a_i,b_1,c_0)\\[1pt]
  \label{eq:Wscale}
  W(q;\nu a_i,\nu b_1,\nu c_0) &=
  \sqrt\nu\,W(\sqrt\nu\,q;a_i,b_1,c_0)\\[1pt]
  \label{eq:cWscale}
  \cW^\pm(q;\nu a_i,\nu b_1,\nu c_0)&=\cW^\pm(\sqrt\nu\,q;a_i,b_1,c_0)\\
  \label{eq:Vscale}
  V^\pm(q;\nu a_i,\nu b_1,\nu c_0) &=
  \nu\,V^\pm(\sqrt\nu\,q;a_i,b_1,c_0)\,. 
\end{align}
\end{subequations}
We shall therefore set $\nu=1$ in the canonical forms 2)--5), the models
corresponding to an arbitrary value of $\nu$ following easily from
Eqs.~\eqref{eq:zscale} and \eqref{eq:EFWV}. It should also be obvious from
Eq.~\eqref{eq:ellint} that
the change of variable $z(q)$, and hence the functions $E$, $F$, $W$ and the
potentials $V^\pm$ determining each model, are defined up to the
transformation $q\mapsto\pm(q-q_0)$, where $q_0\in\bbR$ is a constant. We
shall make use of this observation to simplify the expressions for $E$, $F$,
$W$, and $V^\pm$. 

\smallskip
\nid\textbf{Case 1.}\quad $\D A(z)=2z$.\\[3pt]
\emph{Change of variable}:\quad $\D z=q^2$.\\[3pt]
\emph{Supercharge}:
\begin{equation}
  \label{eq:sc1}
  E = \frac1q\,,\qquad F=\frac2q\,,\qquad W = -\frac{b_1}2\,q
  +\frac{1-\N-2\al^-}{2q}\,. 
\end{equation}
\emph{Potentials}:
\begin{equation}
  \label{eq:pots1}
  V^\pm = \frac18\,b_1^2\,q^2+\frac{\al^\pm(\al^\pm-1)}{2q^2}
  \mp\frac{\N}4 b_1+V_0\,. 
\end{equation}
Here, and in what follows, $V_0$ denotes an arbitrary constant. 

\nid\emph{Solvable sectors}:
\begin{subequations}
  \label{eq:ssec1}
\begin{align}
  \label{eq:ssec1a}
  \cV_{\N_1}^\pm&=\ee^{\mp\frac14 b_1 q^2}q^{\al^\pm}
   \left\langle 1,q^2,\dots,q^{2(\N_1-1)}\right\rangle\,,\\
  \label{eq:ssec1b}
  \cV_{\N_2}^\pm&=\ee^{\mp\frac14 b_1 q^2}q^{1-\al^\pm}
   \left\langle 1,q^2,\dots,q^{2(\N_2-1)}\right\rangle\,,
\end{align}
\end{subequations}
where the linear spaces $\cV_{\N_i}^\pm$ are defined in Eq.~\eqref{eq:cVNpm}. 
This case corresponds to a solvable model.  The potentials \eqref{eq:pots1}
are singular at the origin, and thus their Hamiltonians may be naturally
defined on, e.g., the half-line $S=(0,\infty)$.  In this case, we see from
Eqs.~\eqref{eq:ssec1} that the normalizability of the solvable sectors depends
on the value of $\al^\pm$ (or $\la$) and the sign of $b_1$. The finiteness of
the $L^2$ norm in the solvable sectors yields the following
conditions:\footnote{For the solvable sectors to be included in a Hilbert
  space on which the Hamiltonians $H^\pm$ are \emph{self-adjoint}, we need of
  course to impose a stronger restriction on the parameters coming from the
  boundary condition at the endpoint $q=0$, which is usually taken as
  $\lim_{q\to0+}q^{-1/2}\psi(q)=0$. See also the discussion in Section
  \ref{sec:discus}.} 
\begin{subequations}
  \label{eq:ncon1}
\begin{align}
  \cV_{\N_1}^+\subset L^2(S)\quad &\Lra\quad b_1>0\,,\quad\la>\N_1-\N_2-1\,,\\
  \cV_{\N_2}^+\subset L^2(S)\quad &\Lra\quad b_1>0\,,\quad\la<\N_1-\N_2+1\,,\\
  \cV_{\N_1}^-\subset L^2(S)\quad &\Lra\quad b_1<0\,,\quad\la<1\,,\\
  \cV_{\N_2}^-\subset L^2(S)\quad &\Lra\quad b_1<0\,,\quad\la>-1\,. 
\end{align}
\end{subequations}

\smallskip
\nid\textbf{Case 2a.}\quad $\D A(z)=z^2/2$.\\[3pt]
\emph{Change of variable}:\quad $\D z=\ee^q$.\\[3pt]
\emph{Supercharge}:
\begin{equation}
  \label{eq:sc2}
  E = F = 1\,,\qquad W = -b_1\,. 
\end{equation}
\emph{Potentials}:
\begin{equation}
  \label{eq:pots2}
  V^\pm = V_0\,. 
\end{equation}
Thus both potentials are equal and trivial in this case. 

\smallskip
\nid\textbf{Case 2b.}\quad $\D A(z)=-z^2/2$.\\[3pt]
The formulas for the supercharge and the potentials for this case can be
easily deduced from those of the preceding one by applying
Eqs.~\eqref{eq:EFWV} with $\nu=-1$. 

\smallskip
\nid\textbf{Case 3a.}\quad $\D A(z)=2z(1-z)(1-m\,z)$.\\[3pt]
\emph{Change of variable}:\quad $\D z=\sn^2 q$.\\[3pt]
Here, and in the following cases, the Jacobian elliptic functions have modulus
$k=\sqrt m$. Strictly speaking, if $0<m<1$ the above change of variable is
only valid in one of the two regions in which $A(z)$ is positive, namely the
interval $0<z<1$. In the second region of positivity $1/m<z$ the change of
variable is $z=1/(m\,\sn^2 q)$. However, since the projective transformation
$w=1/(m z)$ leaves $A$ invariant and maps the interval $(0,1)$ into the
half-line $(1/m,\infty)$, we need not consider the second change of variable. 

\smallskip
\nid\emph{Supercharge}:
\begin{equation}
  \label{eq:sc3}
  \begin{gathered}
    E = \frac{3\,m\sn^4 q-2(1+m)\sn^2q+1}{\sn q\,\cn q\,\dn q}\,,\qquad F =
    2\,\frac{\cn q\,\dn q}{\sn q}\,,\\[3pt]
    W = \frac{(1-\N-2\al^-)(1+m \sn^4 q)-b_1\sn^2 q}{2\,\sn q\,\cn q\,\dn
      q}\,. 
  \end{gathered}
\end{equation}
\emph{Potentials}:
\begin{multline}
  \label{eq:pots3}
  V^\pm = \frac m2\,\al^\mp(\al^\mp-1)\sn^2 q
  +\frac{\al^\pm(\al^\pm-1)}{2\sn^2q}
  +\frac{m'}2\,\frac{\be^\pm(\be^\pm-1)}{\cn^2 q}\\
  {}-\frac{m'}2\,\frac{\be^\mp(\be^\mp-1)}{\dn^2 q} +\frac{m'\N}2\,\be^\pm
   +V_0\,. 
\end{multline}
\emph{Parameters}:
\begin{equation}
  \label{eq:bm3}
  \be^\pm=\frac12(1-\N)\pm\frac1{2m'}\bigl(b_1+(1+m)(\N+2\al^--1)\bigr)\,. 
\end{equation}
\emph{Solvable sectors}:
\begin{subequations}
  \label{eq:ssec3}
\begin{align}
  \cV_{\N_1}^\pm&=(\sn q)^{\al^\pm}(\cn q)^{\be^\pm}(\dn q)^{\be^\mp}
   \left\langle 1,\sn^2 q,\dots,(\sn q)^{2(\N_1-1)}\right\rangle\,,\\
  \cV_{\N_2}^\pm&=(\sn q)^{1-\al^\pm}(\cn q)^{\be^\pm}(\dn q)^{\be^\mp}
   \left\langle 1,\sn^2 q,\dots,(\sn q)^{2(\N_2-1)}\right\rangle\,. 
\end{align}
\end{subequations}
It is worth mentioning that in this case the potentials $V^+$ and $V^-$ are
related by a complex translation (up to a constant term), namely
\begin{equation}
  \label{eq:trans3}
  V^+(q)=V^-(q+\ii K')+\frac{\N m'}2\,(\be^--\be^+)\,. 
\end{equation}
In the latter equation $K'\equiv K(m')$ is the complete elliptic integral of
the first kind, defined by
\begin{equation}
  \label{eq:ceii}
  K(m)\equiv\int_0^{\pi/2}\frac{dt}{\sqrt{1-m\sin^2 t}}\,. 
\end{equation}
In general, the potentials \eqref{eq:pots3} have real singularities at the
points $q=nK$ $(n\in\bbZ)$, and thus their Hamiltonians may be naturally
defined on, e.g., $S=(0,K)$.  In this case, we see from Eqs.~\eqref{eq:ssec3}
that the normalizability of the solvable sectors depends on the value of
$\al^\pm$ and $\be^\pm$.  The restrictions on the parameters coming from the
finiteness of the $L^2$ norm in the solvable sectors are now given by
\begin{subequations}
  \label{eq:ncon3}
\begin{alignat}{3}
  \cV_{\N_1}^\pm\subset L^2(S)\quad &\Lra\quad & \al^\pm &>-\frac12\,,
   \quad & \be^\pm &>-\frac12\,,\\
  \cV_{\N_2}^\pm\subset L^2(S)\quad &\Lra & \al^\pm &<\frac32\,,
   & \be^\pm &>-\frac12\,. 
\end{alignat}
\end{subequations}
As previously discussed, when $m=0$ the polynomial
$A(z)$ is of second degree, and hence the Hamiltonians $H^\pm$ are solvable. 
The formulas for the supercharge, the potentials and the solvable sectors
are obtained from Eqs.~\eqref{eq:sc3}--\eqref{eq:ssec3} by setting $m=0$,
$m'=1$, and $(\sn q,\cn q,\dn q)=(\sin q,\cos q,1)$. 

\smallskip
\nid\textbf{Case 3b.}\quad $\D A(z)=-2z(1-z)(1-m\,z)$. 

As in Case 2b, the formulas for this case can follow from those of the
preceding one using Eqs.~\eqref{eq:EFWV} with $\nu=-1$. 
The following well-known identities \cite{GR00} may be of help in this case:
\[
  \sn(\ii q;m) = \ii\,\frac{\sn(q;m')}{\cn(q;m')}\,,\quad
  \cn(\ii q;m) = \frac1{\cn(q;m')}\,,\quad
  \dn(\ii q;m) = \frac{\dn(q;m')}{\cn(q;m')}\,. 
\]
The value $m=0$ yields again solvable models, whose supercharge, potentials
and solvable sectors follow from Eqs.~\eqref{eq:sc3}--\eqref{eq:ssec3} by
setting $m=0$, $m'=1$, and $(\sn\ii q,\cn\ii q,\dn\ii q)=(\ii\sinh q, \cosh
q,1)$. This case deserves further discussion, since the resulting potentials
\begin{align}
  \label{eq:pots3b}
  V^\pm = \frac{\al^\pm(\al^\pm-1)}{2\sinh^2q}
   -\frac{\be^\pm(\be^\pm-1)}{2\cosh^2 q}+\frac12\,\be^\mp(\be^\mp-1)
   -\frac{\N}2\,\be^\pm -V_0\,,
\end{align}
are now of hyperbolic type, and hence are not periodic on $\bbR$. The solvable
sectors are given by
\begin{subequations}
  \label{eq:ssec3b}
\begin{align}
  \cV_{\N_1}^\pm&=(\sinh q)^{\al^\pm}(\cosh q)^{\be^\pm}
   \left\langle 1,\sinh^2 q,\dots,(\sinh q)^{2(\N_1-1)}\right\rangle,\\
  \cV_{\N_2}^\pm&=(\sinh q)^{1-\al^\pm}(\cosh q)^{\be^\pm}
   \left\langle 1,\sinh^2 q,\dots,(\sinh q)^{2(\N_2-1)}\right\rangle. 
\end{align}
\end{subequations}
The potentials \eqref{eq:pots3b} are singular only at the origin, and thus
their Hamiltonians may be naturally defined on, e.g., $S=(0,\infty)$. The
finiteness of the $L^2$ norm in the solvable sectors leads to the following
restrictions:
\begin{subequations}
  \label{eq:ncon3b}
\begin{align}
  \cV_{\N_1}^\pm\subset L^2(S)\quad &\Lra\quad -\frac12<\al^\pm
   <-\be^\pm-2\N_1+2\,,\\
  \cV_{\N_2}^\pm\subset L^2(S)\quad &\Lra\quad \be^\pm +2\N_2-1
   <\al^\pm <\frac32\,. 
\end{align}
\end{subequations}
Note that the above inequalities cannot be satisfied unless $\be^\pm<-2\N_i
+5/2$, $i=1,2$. 

\smallskip
\nid\textbf{Case 4.}\quad $\D A(z)=2z(1-z)(m'+m\,z)$.\\[3pt]
\emph{Change of variable}:\quad $\D z=\cn^2 q$.\\[3pt]
Again, the above change of variable is only valid in the interval $0<z<1$. In
the second region of positivity of $A(z)$, namely the half-line $z<-m'/m$, the
correct change of variable is $z=-m'/(m\,\cn^2 q)$. As before, we shall
restrict ourselves to the interval $0<z<1$, since the projective
transformation $w=-m'/(m z)$ maps this interval into the half-line
$(-\infty,-m'/m)$ and leaves $A$ invariant. 

\smallskip
\nid\emph{Supercharge}:
\begin{equation}
  \label{eq:sc4}
  \begin{gathered}
    E = \frac{3\,m\sn^4 q-2(1+m)\sn^2q+1}{\sn q\,\cn q\,\dn q}\,,\qquad F =
    -2\,\frac{\sn q\,\dn q}{\cn q}\,,\\[3pt]
    W = \frac{(1-\N-2\al^-)(m \cn^4q-m') + b_1\cn^2 q}{2\,\sn q\,\cn q\,\dn
      q}\,. 
  \end{gathered}
\end{equation}
\emph{Potentials}:
\begin{multline}
  \label{eq:pots4}
  V^\pm = \frac m2\,\al^\mp(\al^\mp-1)\sn^2 q
  +\frac{\be^\pm(\be^\pm-1)}{2\sn^2q}
  +\frac{m'}2\,\frac{\al^\pm(\al^\pm-1)}{\cn^2 q}\\
  {}-\frac{m'}2\,\frac{\be^\mp(\be^\mp-1)}{\dn^2 q}
  +\frac m2\,(\N_2-\N_1)\al^\pm+\frac\N2\be^\pm + V_0\,. 
\end{multline}
\emph{Parameters}:
\begin{equation}
  \label{eq:bm4}
  \be^\pm=\frac12(1-\N)\pm\frac12\bigl(b_1+(1-2m)(\N+2\al^--1)\bigr)\,. 
\end{equation}
As in the previous case, the scalar potentials $V^\pm$ are related by a
complex translation:
\begin{equation}
  \label{eq:trans4}
  V^+(q)=V^-(q+K+\ii K')\,. 
\end{equation}\emph{Solvable sectors}:
\begin{subequations}
  \label{eq:ssec4}
\begin{align}
  \cV_{\N_1}^\pm&=(\cn q)^{\al^\pm}(\sn q)^{\be^\pm}(\dn q)^{\be^\mp}
   \left\langle 1,\cn^2 q,\dots,(\cn q)^{2(\N_1-1)}\right\rangle,\\
  \cV_{\N_2}^\pm&=(\cn q)^{1-\al^\pm}(\sn q)^{\be^\pm}(\dn q)^{\be^\mp}
   \left\langle 1,\cn^2 q,\dots,(\cn q)^{2(\N_2-1)}\right\rangle. 
\end{align}
\end{subequations}
The potentials are again singular at integer multiples of $K$, so that
their Hamiltonians $H^\pm$ can be taken as defined on, e.g., $S=(0,K)$. 
The conditions ensuring the finiteness of the $L^2$ norm in the solvable
sectors are given by Eqs.~\eqref{eq:ncon3}. 

\smallskip \nid\textbf{Case 5.}\quad $\D A(z)=
\frac12\,z\bigl(z^2+2(1-2m)z+1\bigr)$.\\[3pt]
\emph{Change of variable}:\quad $\D z=\frac{1+\cn q}{1-\cn q}$.\\[3pt]
\smallskip \emph{Supercharge}:
\begin{equation}
  \label{eq:sc5}
  \begin{gathered}
    E =- \frac{\cn q+2\dn^2q}{\sn q\,\dn q}\,,\qquad F =
    -2\,\frac{\dn q}{\sn q}\,,\\[3pt]
    W = \frac{(2b_1-\N-2\al^-+1)\sn^2q+2(\N+2\al^--1)}{4\,\sn q\,\dn q}\,. 
  \end{gathered}
\end{equation}
\emph{Potentials}:
\begin{equation}
  \label{eq:pots5}
  V^\pm = \frac{\al^\pm(\al^\pm-1)}{4(1+\cn q)}
  +\frac{\al^\mp(\al^\mp-1)}{4(1-\cn q)}
  +\frac{\N m\be^\pm\cn q+(\be^\pm)^2+\frac14\,mm'(1-\N^2)}{2\dn^2q} + V_0\,. 
\end{equation}
\emph{Parameters}:
\begin{equation}
  \label{eq:bm5}
  \be^\pm=\pm\frac14\bigl(2b_1-(1-2m)(\N+2\al^--1)\bigr)\,. 
\end{equation}
\par\noindent
The scalar potentials $V^\pm$ are related by a \emph{real}
translation, namely
\begin{equation}
  \label{eq:trans5}
  V^+(q)=V^-(q+2K)\,. 
\end{equation}
Hence $H^+$ and $H^-$ are \emph{self-isospectral} in this case,
cf.~\cite{DF98}. 

\noindent\emph{Gauge factors}:
\begin{equation}
  \label{eq:gauge5}
  \ee^{-\cU^\pm}= \Big(\frac{1-\cn q}{\dn q}\Big)^{\frac{\N-1}2}
   \exp\left(-\frac{\be^\pm}{k k'}\, \arctan\frac{k^2 \cn q+k'^2}{
   kk'(1-\cn q)}\right),
\end{equation}
where $k'=\sqrt{m'}\equiv\sqrt{1-m}$. 

The potentials $V^\pm$ are singular at integer
multiples of $2K$, so that their corresponding Hamiltonians are naturally
defined on, e.g., the interval $S=(0,2K)$. 
The necessary and sufficient conditions ensuring the square integrability of
the wave functions in the solvable sectors are given by
\begin{subequations}
  \label{eq:ncon5}
\begin{align}
  \label{eq:ncon5a}
  \cV_{\N_1}^\pm\subset L^2(S)\quad &\Lra\quad -\frac12<\al^\pm
   <\N_2-\N_1+\frac32\,,\\
   \label{eq:ncon5b}
  \cV_{\N_2}^\pm\subset L^2(S)\quad &\Lra\quad -\frac12 +\N_2-\N_1
   <\al^\pm <\frac32\,. 
\end{align}
\end{subequations}
Note, in particular, that the condition $\N_1\geq\N_2$ implies that the
inequalities \eqref{eq:ncon5a} cannot hold unless $\N_1-\N_2=0,1$. 

\section{$\N$-fold superalgebra and associated polynomial families}
\label{sec:BDpoly}
In ordinary (``one-fold'') supersymmetric quantum mechanics, it is well known
that the superHamiltonian $\bH$ is proportional to the anticommutator of the
supercharges $Q_\N^\pm$, and hence the operators $Q_\N^\pm$ and $\bH$ span a
three-dimensional Lie superalgebra. This cannot possibly be the case in
$\N$-fold supersymmetric quantum mechanics (with $\N>1$), since the
anticommutator
\begin{equation}
  \label{eq:motherH}
  \cH_\N\equiv\frac12\,\bigl\{Q_\N^+,Q_\N^-\bigr\}
\end{equation}
is now a linear differential operator of order $2\N$. Note, however, that from
the nilpotency of the supercharges (Eq.~\eqref{eq:QmQp}) and the
supersymmetric character of $\bH$ (Eq.~\eqref{eq:QpmH}), it immediately
follows that the anticommutator \eqref{eq:motherH}, that we shall henceforth
call the \emph{mother Hamiltonian}, commutes with $Q_\N^\pm$
and with $\bH$, namely
\begin{equation}
  \label{eq:mHHQ}
  \bigl[\cH_\N,Q_\N^\pm\bigr]=\bigl[\cH_\N,\bH\bigr]=0\,. 
\end{equation}
These relations strongly suggest that the mother Hamiltonian $\cH_\N$ is a
polynomial of degree $\N$ in $\bH$, say $\cH_\N=\Pi_\N(\bH)$, and hence that
the operators $Q_\N^\pm$ and $\bH$ span a nonlinear superalgebra of
degree $\N$ defined by the relations \eqref{eq:QmQp}, \eqref{eq:QpmH}, and
\begin{equation}
  \label{eq:PiN}
  \bigl\{Q_\N^+,Q_\N^-\bigr\}=2\,\Pi_\N(\bH)\,. 
\end{equation}
That this is indeed the case was rigorously proved in Refs.~\cite{AST01b,%
AS03}, where it was also shown that $\Pi_\N$ is proportional to the
characteristic polynomial of the restriction of the component Hamiltonians
$H^\pm$ to the invariant spaces $\cV_\N^\pm$, namely
\begin{equation}
  \label{eq:mHcharpoly}
  \Pi_\N(E)=2^{\N-1}\det\bigl(H^\pm|_{\cV_\N^\pm}-E\bigr)\,. 
\end{equation}
By Eqs.~\eqref{eq:Hpmgauge} and \eqref{eq:modules}, this is equivalent to the
``gauged'' relation
\begin{equation}
  \label{eq:mHgauged}
  \Pi_\N(E)=2^{\N-1}\det\bigl(\bar{\tH}^\pm|_{\bar{\tcV}_\N^\pm}-E\bigr)\,. 
\end{equation}
It was later shown in Ref.~\cite{Ta03a} that for type A models the right-hand
side of the latter equation is proportional to the critical generalized
Bender--Dunne polynomial (GBDP) associated to the gauged Hamiltonian
$\bar{\tH}^\pm$, which in this case is Lie-algebraic with respect to the Lie
algebra $\fsl(2)$. Bender--Dunne polynomials were introduced in
Ref.~\cite{BD96} to determine the
solvable part of the spectrum of a well-known quasi-exactly solvable sextic
oscillator Hamiltonian \cite{TU87}, and were soon generalized by Finkel et
al.~\cite{FGR96} to virtually all one-dimensional quasi-exactly solvable
models associated to the $\fsl(2)$ algebra. 

We shall now determine the polynomial $\Pi_\N$ for the type C models
constructed in the previous section; in particular, we shall show that in this
case $\Pi_\N$ factorizes (up to a multiplicative constant) as the product of
two critical Bender--Dunne type polynomials of degrees $\N_1$ and $\N_2$.  In
view of the equality of the characteristic polynomials of the operators
$\tH^-|_{\tcV_\N^-}$ and $\Hbp |_{\bcVp_\N}$ (which follows in general from
Eq.~\eqref{eq:mHgauged}), it suffices to study the action of one of the gauged
Hamiltonians $\,\bar{\tH}^\pm$ in its corresponding invariant space, e.g., of
$\,\tH\equiv\tH{}^-$ in $\tcV_\N\equiv\tcV_\N^-$. 

Recall, to begin with, that the linear space $\tcV_\N$ is the direct sum
\begin{equation}
  \label{eq:tcVNstruct}
  \tcV_\N=\bigl\langle 1,z,\dots,z^{\N_1-1}\bigr\rangle \oplus
  z^\la\,\bigl\langle 1,z,\dots,z^{\N_2-1}\bigr\rangle
  \equiv\tcV^{(\text A)}_{\N_1}\oplus \tcV^{(\text A)}_{\N_2}
\end{equation}
of the two spaces $\tcV^{(\text A)}_{\N_i}$ (cf.~Eq.~\eqref{eq:typeA}),
each of which is invariant under the action of $\,\tH$. Hence
\begin{equation}
  \label{eq:factCP}
  \det\bigl(\tH|_{\tcV_\N}-E\bigr)=\det\bigl(\tH^{(1)}-E\bigr)
  \cdot\det\bigl(\tH^{(2)}-E\bigr)\,,
\end{equation}
where we have set
\begin{equation}
  \label{eq:Hi}
  \tH^{(i)}=\tH\raisebox{-2pt}{$\bigr|$}\vphantom{p}_{\tcV^{\text{(A)}}_{\N_i}}. 
\end{equation}
The second-order operator $z^{-(i-1)\la}\,\tH^{(i)}z^{(i-1)\la}$ ($i=1,2$)
preserves a type A space \eqref{eq:typeA} of dimension $\N_i$, and the
coefficient of $\pa_z^2$ in this operator is the polynomial $A(z)$. Since
$A(z)$ has degree three and vanishes at the origin (see Eq.~\eqref{eq:az}),
it follows \cite{FGR96} that each of the operators
$z^{-(i-1)\la}\,\tH^{(i)}z^{(i-1)\la}$ ($i=1,2$) defines an associated family
of GBDPs $\{\pi^{[\N_1,\N_2]}_{i,k}\}_{k=0}^\infty$ whose critical element
$\pi^{[\N_1,\N_2]}_{i,\N_i}$ (since $\N_i=\dim\tcV^{\text{(A)}}_{\N_i}$) is
proportional to the characteristic polynomial of
$z^{-(i-1)\la}\,\tH^{(i)}z^{(i-1)\la}$, and hence of $\,\tH^{(i)}$:
\begin{equation}
  \label{eq:critBDi}
  \det\bigl(\tH^{(i)}-E\bigr)=\det\bigl(z^{-(i-1)\la}\,\tH^{(i)}z^{(i-1)\la}
   -E\bigr)
  =(-1)^{\N_i}\pi^{[\N_1,\N_2]}_{i,\N_i}(E)\,,
  \qquad i=1,2\,. 
\end{equation}
By Eqs.~\eqref{eq:mHgauged} and \eqref{eq:factCP} we thus have
\begin{equation}
  \label{eq:PiNpis}
  \Pi_\N = (-1)^\N2^{\N-1}\pi^{[\N_1,\N_2]}_{1,\N_1}\,
\pi^{[\N_1,\N_2]}_{2,\N_2}\,,
\end{equation}
and consequently (cf.~Eq.~\eqref{eq:PiN}) the type C $\N$-fold superalgebra is
given by
\begin{subequations}
\begin{align}
  \label{eq:SUSYalg}
  &\bigl\{Q_\N^\pm,Q_\N^\pm\bigr\}=\bigl[Q_\N^\pm,\bH\bigr]=0\,,\\
  \label{eq:SUSYalgb}
  &\bigl\{Q^+_\N,Q^-_\N\bigr\}=(-2)^\N\pi^{[\N_1,\N_2]}_{1,\N_1}(\bH)\,
  \pi^{[\N_1,\N_2]}_{2,\N_2}(\bH)\,. 
\end{align}
\end{subequations}
We should note at this point that, as previously remarked, the operators of
the form $z^{-(i-1)\la}\,\tH^{(i)}z^{(i-1)\la}$ ($i=1,2$) do not exhaust
\emph{all} the possible gauged Hamiltonians of type A. As a consequence, some
of the characteristic features of the type A models are inevitably lost in the
type C case. In particular, the $\mathrm{GL}(2,\bbR)$ invariance of type A
models, which ensures that all the coefficients of the associated GBDPs are
expressed in terms of polynomial invariants~\cite{Ta03a}, is broken in type C
models and, as a consequence, the same is true for the $\N$-fold superalgebra
of type C. In other words, the type C models have two invariant subspaces of
type A at the cost of the $\mathrm{GL}(2,\bbR)$ symmetry. On the other hand,
it turns out that each of the polynomial families associated with type C
models acquires a novel feature, namely the dependency on \emph{two} positive
integers $\N_1$ and $\N_2$ (cf.~Eq.~\eqref{eq:pirr2} below). That is the
reason for the rather cumbersome notation $\pi_{i,k}^{[\N_1,\N_2]}$, that we
shall hereafter abbreviate as $\pi_{i,k}$ unless the dependence on $\N_1$ and
$\N_2$ is crucial. 

To construct the polynomial families $\{\pi_{i_k}\}_{k=0}^\infty$ ($i=1,2$)
associated with type C models, let $\chi_E(z)$ denote an eigenfunction of $\,\tH$
with eigenvalue $E$. In view of Eq.~\eqref{eq:tcVNstruct}, we shall consider
the following two formal expansions of this eigenfunction in powers of $z$:
\begin{equation}
  \label{eq:chiEexp}
  \chi_E(z) = z^{(i-1)\la}\,\sum_{k=0}^\infty\,
  \frac{\hpi_{i,k}(E)}{\Ga(k+1+(i-1)\la)}\:z^k\,,
  \qquad i=1,2\,. 
\end{equation}
Here we have set $\hpi_{i,k}=\ga_{i,k}\,\pi_{i,k}$, $\ga_{i,k}$ being
a numerical coefficient that must be chosen so that $\pi_{i,k}$ is monic. 
Clearly, the necessary and sufficient condition for $\chi_E(z)$ to belong
to $\tcV^{\text{(A)}}_{\N_i}$
is that
\begin{equation}
  \label{eq:eigcond}
  \hpi_{i,k}(E)=0\,,\qquad \text{for all }k\geq\N_i\,. 
\end{equation}
Acting on $\chi_E(z)$ with $\,\tH$ and using Eqs.~\eqref{eq:tH},
\eqref{eq:abcz}, and \eqref{eq:b1} one immediately arrives at the following
recursion relation for the coefficients $\hpi_{i,k}(E)$:
\begin{multline}
  \label{eq:hpirr}
   a_1\bigl(k+(i-2)\la+1\bigr)\hpi_{i,k+1}={}\\
    -\Bigl[E+c_0+\bigl(k+(i-1)\la\bigr)\bigl(b_1
   +a_2\bigl(k+(i-1)\la-\N+1\bigr)\bigr)\Bigr]\hpi_{i,k}\\
    -a_3\bigl(k+(i-1)\la\bigr)
   \Bigl[\bigl(k+(i-1)\la-1\bigr)\bigl(k+(i-2)\la-\N+1\bigr)\\
    +(\N_1-1)(\N_2+\la-1)\Bigr]\hpi_{i,k-1}\,;\qquad
   k\geq0\,,\quad\hpi_{i,-1}\equiv0\,. 
\end{multline}
Note that the coefficient $a_1$ is nonzero in all the canonical forms listed
in Table \ref{tb:table1} with the exception of the second one, which
corresponds to the trivial case of a constant potential. Hence we shall assume
in what follows that $a_1\neq0$. In that case the recurrence relation
\eqref{eq:hpirr} can be brought to the more standard form
\begin{multline}
  \label{eq:pirr}
  \pi_{i,k+1}=\Bigl[E+c_0+\bigl(k+(i-1)\la\bigr)\bigl(b_1
   +a_2\bigl(k+(i-1)\la-\N+1\bigr)\bigr)\Bigr]\pi_{i,k}\\
  -a_1\,a_3\bigl(k+(i-1)\la\bigr)\bigl(k+(i-2)\la\bigr)
   \Bigl[\bigl(k+(i-1)\la-1\bigr)\\
  \times\bigl(k+(i-2)\la-\N+1\bigr)+(\N_1-1)(\N_2+\la-1)
   \Bigr]\pi_{i,k-1}\,;\\
  \qquad k\geq0\,,\quad\pi_{i,0}\equiv1\,,
\end{multline}
by choosing the so far undetermined multipliers $\ga_{i,k}$ as follows:
\begin{equation}
\label{eq:gammaik}
\ga_{i,k}=\frac1{(-a_1)^k\,\Ga(k+1+(i-2)\la)}\,,\qquad k\geq0\,. 
\end{equation}
The three-term recursion relation \eqref{eq:pirr} will actually define a
family of \emph{weakly} orthogonal polynomials
$\{\pi_{i,k}\}_{k=0}^\infty$ provided that the coefficient of $\pi_{i,k-1}$
vanishes for some non-negative integer value of $k$ \cite{Ch78}. If $k=K$ is
the lowest such value, and $E_j$ ($j=1,\dots,K$) is a root of the
\emph{critical polynomial} $\pi_{i,K}$, then the recursion relation
\eqref{eq:pirr} implies that $\pi_{i,k}(E_j)=0$ for all $k\geq K$. It
follows~\cite{BD96,FGR96}
that the linear space $z^{(i-1)\la}\langle 1,z,\dots,z^{K-1}\rangle$ is
invariant under $\,\tH$, and that the eigenvalues of the
restriction of $\,\tH$ to this space are the $K$ roots (counting
multiplicities) of the critical polynomial $\pi_{i,K}$.  In particular, the
characteristic polynomial of the restriction of $\,\tH$ to the invariant space
$z^{(i-1)\la}\langle 1,z,\dots,z^{K-1}\rangle$ is proportional to
$\pi_{i,K}(E)$. For the recursion relation \eqref{eq:pirr}, the coefficient of
$\pi_{i,k-1}$ can be written as
\begin{equation}
  \label{eq:critcoeff}
  \begin{cases}
    -a_1 a_3 k(k-\N_1)(k-\la)(k-\N_2-\la)\,,\quad & i=1\\[2mm]
    -a_1 a_3 k(k-\N_2)(k+\la)(k-\N_1+\la)\,,\quad & i=2\,. 
  \end{cases}
\end{equation}
Taking into account the restrictions \eqref{eq:restla} on $\la$, it is easily
seen that the degree of the critical polynomial is $K=\N_i$, $i=1,2$. This
establishes Eq.~\eqref{eq:critBDi}, since the space $z^{(i-1)\la}\langle
1,z,\dots,z^{\N_i-1}\rangle$ coincides with $\tcV^{\text{(A)}}_{\N_i}$ ($i=1,2$) by
Eq.~\eqref{eq:tcVNstruct}. 

It is worth mentioning that the polynomial systems $\pi_{i,k}$ associated with
type C $\N$-fold supersymmetry are always weakly orthogonal in spite of the
fact that the solvable sectors $\cV_\N^\pm$ are not always normalizable, as was
studied in the preceding section.  We thus obtain further evidence of the
claim in Ref.~\cite{Ta03a} that normalizability has in general nothing to do
with the weak orthogonality of the associated GBDPs.  Note also that
Eq.~\eqref{eq:pirr} becomes a two-term recursion relation if and only if the
coefficient $a_3$ vanishes, or, equivalently, if the corresponding Hamiltonian
is solvable in Turbiner's sense. 

The recursion relation \eqref{eq:pirr} determining the Bender--Dunne type
polynomials $\pi_{i,k}$ and, ultimately, the $\N$-fold superalgebra via
Eq.~\eqref{eq:SUSYalgb}, can be recast into the following more concise form
using Eq.~\eqref{eq:critcoeff}:
\begin{align}
\label{eq:pirr2}
 \lefteqn{
  \pi_{i,k+1}=\Bigl[E+c_0+\bigl(k+(i-1)\la\bigr)\bigl(b_1
   +a_2\bigl(k+(i-1)\la-\N+1\bigr)\bigr)\Bigr]\pi_{i,k}
 }\notag\\
  &{}-a_1a_3\,k(k-\N_i)\bigl(k+(2i-3)\la\bigr)\bigl(k-\N_{3-i}+(2i-3)\la
   \bigr)\pi_{i,k-1}\,. 
\end{align}
This recursion relation can be used without difficulty to compute the
polynomial families $\pi_{i,k}^{[\N_1,\N_2]}$ and, in particular, the critical
polynomials determining the $\N$-fold superalgebra, for any given values of
$\N_1$ and $\N_2$. We shall exhibit in what follows a few examples of these
polynomials for $\N=3$ and $4$, assuming that $\N_1$ and $\N_2$ satisfy the
restriction $\N_1\geq\N_2$ imposed in Section~\ref{sec:constr}. 

\smallskip 
\nid\textbf{Example 1.}\quad $\N_1=2,\,\N_2=1$.\\[6pt]
\emph{Polynomial system}:
\begin{subequations}
  \begin{align}
    \pi_{1,1}^{[2,1]}(E) &= E+c_0\,,\\
    \pi_{1,2}^{[2,1]}(E) &= (E+c_0)(E+c_0+b_1-a_2)
     +\la(\la-1)a_1 a_3\,,\\[5pt]
    \pi_{2,1}^{[2,1]}(E) &= E+c_0+\la b_1 +\la(\la-2)a_2\,. 
  \end{align}
\end{subequations}
\emph{3-fold superalgebra}:
\begin{equation}
  \bigl\{Q_3^+,Q_3^-\bigr\}= -8\bigl[(\bH+c_0)(\bH+c_0
   +b_1-a_2)+\la(\la-1)a_1 a_3\bigr]
  \bigl[\bH+c_0+\la b_1+\la(\la-2)a_2\bigr]. 
\end{equation}

\smallskip \nid\textbf{Example 2.}\quad $\N_1=3,\,\N_2=1$.\\[6pt]
\emph{Polynomial system}:
\begin{subequations}
  \begin{align}
    \pi_{1,1}^{[3,1]}(E)=&\, E+c_0\,,\\
    \pi_{1,2}^{[3,1]}(E)=&\, (E+c_0)(E+c_0+b_1-2a_2)
     +2\la(\la-1)a_1 a_3\,,\\
    \pi_{1,3}^{[3,1]}(E)=&\, (E+c_0)(E+c_0+b_1-2a_2)(E+c_0+2b_1-2a_2)
     \notag\\
     &\, +4(\la-1)a_1 a_3\bigl((\la-1)(E+c_0)+\la(b_1-a_2)\bigr)\\[5pt]
    \pi_{2,1}^{[3,1]}(E)=&\, E+c_0+\la b_1 +\la(\la-3)a_2\,. 
  \end{align}
\end{subequations}
\emph{4-fold superalgebra}:
 \begin{align}
   \bigl\{Q_4^+,Q_4^-\bigr\}=&\, 16\Bigl[(\bH+c_0)(\bH+c_0
    +b_1-2a_2)(\bH+c_0+2b_1-2a_2)\notag\\
   &\, {}+4(\la-1)a_1 a_3\bigl((\la-1)(\bH+c_0)+\la(b_1-a_2)\bigr)\Bigr]
    \notag\\
   &\, {}\times\bigl[\bH+c_0+\la b_1+\la(\la-3)a_2\bigr]\,. 
 \end{align}

\smallskip \nid\textbf{Example 3.}\quad $\N_1=2,\,\N_2=2$.\\[6pt]
\emph{Polynomial system}:
\begin{subequations}
  \begin{align}
    \pi_{1,1}^{[2,2]}(E)=&\, E+c_0\,,\\
    \pi_{1,2}^{[2,2]}(E)=&\, (E+c_0)(E+c_0+b_1-2a_2)
     +(\la^2-1)a_1 a_3\,,\\[5pt]
    \pi_{2,1}^{[2,2]}(E)=&\, E+c_0+\la b_1 +\la(\la-3)a_2\,.\\
    \pi_{2,2}^{[2,2]}(E)=&\, \bigl(E+c_0+(\la+1)b_1
     +(\la+1)(\la-2)a_2\bigr)\notag\\
     &\, {}\times\bigl(E+c_0+\la b_1+\la(\la-3)a_2\bigr)+(\la^2-1)a_1 a_3\,. 
  \end{align}
\end{subequations}
\emph{4-fold superalgebra}:
 \begin{align}
   \bigl\{Q_4^+,Q_4^-\bigr\}=&\, 16\bigl[(\bH+c_0)(\bH+c_0
    +b_1-2a_2)+(\la^2-1)a_1 a_3\bigr]\notag\\
   &\, {}\times\Bigl[\bigl(\bH+c_0+(\la+1)b_1+(\la+1)(\la-2)a_2\bigr)\notag\\
   &\, {}\times\bigl(\bH+c_0+\la b_1+\la(\la-3)a_2\bigr)
    +(\la^2-1)a_1 a_3\Bigr]\,. 
 \end{align}
\section{Summary and discussion}
\label{sec:discus}
In this article we develop an algorithmic procedure for constructing an
$\N$-fold supersymmetric quantum system starting from a given
finite-dimensional space of functions invariant under a suitable gauge
transform of one of the supercharges. Although the method is very general, it
is especially useful when one knows a particular example of quasi-solvable
Hamiltonian and its algebraically solvable wave functions. We have applied
this procedure to the monomial spaces of Post--Turbiner type, thus obtaining
the new models of type C as well as recovering the previously known type A and
type B systems. We would also like to stress that, although the procedure
always yields an $\N$-fold supersymmetric system $(H^\pm,P_\N)$, it does not
rule out the existence of more general systems of the same type. For instance,
for the type C models specifically discussed in this paper the supercharges
are given by Eqs.~\eqref{eq:dfNsc} and \eqref{eq:PNC}, and the method
developed in this paper guarantees that the supersymmetry algebra holds for
suitable Hamiltonians $H^\pm$ provided that  the functions $E$ and $F$ satisfy
Eq.~\eqref{eq:EFid}. It is not clear, however, whether this sufficient
condition is also necessary. It should also be noted in this respect that the
situation is completely analogous for the type B models discussed in
Ref.~\cite{GT03}. 

The normalizability of the solvable sectors of the type C models, which plays
an important role for the existence of dynamical $\N$-fold supersymmetry
breaking~\cite{AST01b,Ta03c}, is briefly investigated in
Section~\ref{sec:class}. {}From Eqs.~\eqref{eq:ncon1}, \eqref{eq:ncon3},
\eqref{eq:ncon3b}, and \eqref{eq:ncon5}, we see that there is little chance
for both sectors $\cV_{\N_1}^+$ and $\cV_{\N_2}^+$ (resp.~$\cV_{\N_1}^-$ and
$\cV_{\N_2}^-$) to be \emph{simultaneously} normalizable. This means that, in
general, only a part of the whole solvable sector $\cV_\N^+$
(resp.~$\cV_\N^-$) is physical in type C $\N$-fold supersymmetric systems. For
example, among the four subsectors $\cV_{\N_i}^\pm$ in Case 1 only
$\cV_{\N_2}^-$ is normalizable and thus physical (boundary conditions aside)
if $b_1<0$ and $\la\geq 1$. We can thus say that $\N$-fold supersymmetry is
\emph{partially} broken in type C models.  This phenomenon is novel for type C
due to the characteristic structure of its solvable sector, and, to the best
of the authors' knowledge, has not been previously mentioned in the
literature.  For a more precise discussion, we need of course to deal with the
boundary conditions at the singularities in order to define a self-adjoint
extension of the Hamiltonians. This kind of mathematical subtlety is beyond
the scope of the present article, and we shall therefore content ourselves
with referring the reader to, e.g, Ref.~\cite{TFC03} for a recent discussion
of this topic. We also note, in this connection, that the significance of the
boundary conditions in (ordinary) supersymmetry breaking was recently reported
in a different context, namely, through the careful calculation of the fermion
determinant arising out of the path integral formalism \cite{Ki03}. 

The structure of the solvable sectors in type C models gives rise to another
interesting phenomenon in the theory of exact solutions of the Schr\"odinger
equation. Indeed, when the parameter $\al^+$ (resp.~$\al^-$) is a positive
integer $l+1$ (in both cases $\la$ is a half-integer,
cf.~Eq.~\eqref{eq:lapm}), the potential $V^+$ (resp.~$V^-$) in Case 1 given by
Eq.~\eqref{eq:pots1} is nothing but the radial harmonic oscillator plus the
centrifugal potential with a \emph{properly} quantized angular momentum $l$. 
It is then apparent that for $b_1>0$ (resp.~$b_1<0$) the first solvable sector
$\cV_{\N_1}^+$ (resp.~$\cV_{\N_1}^-$) ($\N_1=1,2,\dots$) in
Eq.~\eqref{eq:ssec1a} contains the physical solutions around $q=0$, while the
second sector $\cV_{\N_2}^+$ (resp.~$\cV_{\N_2}^-$) in Eq.~\eqref{eq:ssec1b}
yields the ``second'' solutions, usually discarded as unphysical due to their
singularity\footnote{Or their non-vanishing, for $l=0$.} at the origin. 

By Eq.~\eqref{eq:cVNpm}, we  may in fact say that type C models are
characterized by the fact that \emph{both} linearly independent solutions
around $z(q)=0$ are quasi-solvable.  More precisely, from
Eqs.~\eqref{eq:Hgpmcoll}, \eqref{eq:cz}, \eqref{eq:Q-C}, \eqref{eq:twgen}, and
Table \ref{tb:table1}, it is easily seen that the operator $\tH^-_\N-E$
(resp.~$\Hbp_\N-E$) has a regular singularity at $z=0$, so that Fuchs's
theorem applies. Except in the trivial Case 2, which we shall henceforth
ignore, for all values of the energy $E$ the roots of the indicial equation
are $0$ and $\la$ (resp.~$\N_2$ and $\N_2+\bar\la$). From Eqs.~\eqref{eq:csum}
for the space $\tilde\cV_\N^-$, and the analogous decomposition
\[
\bar\cV_\N^+= z^{\N_2}\,\big\langle1,\dots,z^{\N_1-1}\big\rangle
\oplus z^{\N_2+\bar\la}\,\big\langle 1,\dots,z^{\N_2-1}\big\rangle
\]
for $\bar\cV_\N^+$, we see that each of the four sectors of the invariant
spaces $\bar{\tilde\cV}_\N^\pm$ consists of the solvable eigenfunctions of
$\bar{\tilde H}^\pm_\N$ behaving as $z^i$ near $z=0$, where $i$ is one of the
roots of the indicial equation listed above. Note, in this respect, that this
remark is still valid when the difference of the roots of the indicial
equation, which is given by $\la$ (resp.~$\bar\la$), is an integer, even if in
this case Fuchs's theorem can only guarantee the existence of one linearly
independent power series solution around $z=0$. 

The construction of the Bender--Dunne type polynomial systems associated with
the type C models turned out to be straightforward. The breakdown of the
$\mathrm{GL}(2,\bbR)$ symmetry in type C models spoils the characteristic
feature possessed by the type A polynomials. Instead, the GBDPs of type C have
a novel dependence on two positive-integer parameters $\N_1$ and $\N_2$.  For
any given pair $(\N_1,\N_2)\in\bbN\times\bbN$, we obtain two related families of
weakly orthogonal polynomials satisfying the recursion relation
\eqref{eq:pirr2}. The assertion in Ref.~\cite{Ta03a} that normalizability has
nothing to do with the weak orthogonality of the associated family of
polynomials has also been confirmed.  We also note that a similar construction
for the type B systems presents considerable difficulties due to the lack of
form-invariance under projective transformations of these models. This
difficulty has also prevented the systematic calculation of the explicit form
of the type B $\N$-fold superalgebra.

All the quasi-solvable second-order differential operators in one variable
preserving a finite-dimensional linear space of monomials, which were
classified in Ref.~\cite{PT95}, have now been brought into the framework of
$\N$-fold supersymmetry.\footnote{With only a few exceptions for $\N=3$ and
  $\N=4$.}  The natural continuation of the present work is the study of
operators possessing non-monomial type invariant subspaces, and the
construction of the associated supersymmetric models following the general
algorithm developed in Section~\ref{sec:NQES}.  It would also be of great
interest to extend the results obtained in this paper to a multi-dimensional
space-time. Indeed, the generalization of $\N$-fold supersymmetry to
higher-dimensional  space-times remains one of the most challenging open
problems in this field. Recently, Smilga showed \cite{Sm03} that some
$\N$-fold supersymmetric systems with $\N=2$ can be realized as \emph{weakly}
supersymmetric field theories in one space-time dimension. Although the
results in Ref.~\cite{Sm03} clearly indicate the existence of rather severe
obstructions, further investigation in this direction would still be certainly
worth undertaking. 

\section*{Acknowledgments}
  This work was partially supported by Spain's DGI under grant
  no.~BFM2002--02646 (A. G.-L.), as well as by a Spanish Ministry of
  Education, Culture and Sports research fellowship (T. T.).

\bibliography{refs}

\begin{thebibliography}{10}
\providecommand{\url}[1]{\texttt{#1}}
\providecommand{\urlprefix}{URL }
\providecommand{\eprint}[2][]{\url{#2}}

\bibitem{AST01b}
Aoyama H, Sato M and Tanaka T 2001 {$\mathcal N$}-fold supersymmetry in quantum
  mechanics: general formalism \emph{Nucl. Phys. B} \textbf{619} 105--127
  \eprint{quant-ph/0106037}

\bibitem{AS03}
Andrianov A and Sokolov A 2003 Nonlinear supersymmetry in quantum mechanics:
  algebraic properties and differential representations \emph{Nucl. Phys. B}
  \textbf{660} 25--50 \eprint{hep-th/0301062}

\bibitem{Da1882}
Darboux G 1882 Sur une proposition relative aux {\'e}quations lin{\'e}aires
  \emph{Comptes Rendus Acad. Sci.} \textbf{94} 1456--1459

\bibitem{MS91}
Matveev V and Salle M 1991 \emph{{D}arboux {T}ransformations and {S}olitons}
  (Berlin: Springer Verlag)

\bibitem{AIS93}
Andrianov A, Ioffe M and Spiridonov V 1993 Higher-derivative supersymmetry and
  the {W}itten index \emph{Phys. Lett. A} \textbf{174} 273--279
  \eprint{hep-th/9303005}

\bibitem{TU87}
Turbiner A and Ushveridze A 1987 Spectral singularities and quasi-exactly
  solvable quantal problem \emph{Phys. Lett. A} \textbf{126} 181--183

\bibitem{Us94}
Ushveridze A 1994 \emph{{Q}uasi-exactly {S}olvable {M}odels in {Q}uantum
  {M}echanics} (Bristol: IOP Publishing)

\bibitem{RS88}
Rubakov V and Spiridonov V 1988 Parasupersymmetric quantum mechanics \emph{Mod.
  Phys. Lett. A} \textbf{3} 1337--1347

\bibitem{BD90}
Beckers J and Debergh N 1990 Parastatistics and supersymmetry in quantum
  mechanics \emph{Nucl. Phys. B} \textbf{340} 767--776

\bibitem{Kh93}
Khare A 1993 Parasupersymmetry in quantum mechanics \emph{J. Math. Phys.}
  \textbf{34} 1277--1294

\bibitem{Du93a}
Durand S 1993 Fractional supersymmetry and quantum mechanics \emph{Phys. Lett.
  B} \textbf{312} 115--120 \eprint{hep-th/9305128}

\bibitem{Ta03c}
Tanaka T 2003 {$\mathcal N$}-fold supersymmetry and quasi-solvability
  (preprint), to appear in \emph{Progress in Mathematical Physics Research}
  (New York: Nova Science Publishers)

\bibitem{AST01a}
Aoyama H, Sato M and Tanaka T 2001 General forms of a {$\mathcal N$}-fold
  supersymmetric family \emph{Phys. Lett. B} \textbf{503} 423--429
  \eprint{quant-ph/0012065}

\bibitem{Ta03a}
Tanaka T 2003 Type {A} {$\mathcal N$}-fold supersymmetry and generalized
  {B}ender--{D}unne polynomials \emph{Nucl. Phys. B} \textbf{662} 413--446
  \eprint{hep-th/0212276}

\bibitem{GT03}
Gonz\'{a}lez-L\'{o}pez A and Tanaka T 2004 A new family of {$\mathcal N$}-fold
  supersymmetry: type {B} \emph{Phys. Lett. B} \textbf{586} 117--124
  \eprint{hep-th/0307094}

\bibitem{PT95}
Post G and Turbiner A 1995 Classification of linear differential operators with
  invariant subspace in monomials \emph{Russ. J. Math. Phys.} \textbf{3}
  113--122 \eprint{funct-an/9307001}

\bibitem{Ta04}
Tanaka T 2004 {$\fsl({M}+1)$} construction of quasi-solvable quantum {$M$}-body
  systems \emph{Ann. Phys.} \textbf{309} 239--280 \eprint{hep-th/0306174}

\bibitem{Sh89}
Shifman M 1989 New findings in quantum mechanics (partial algebraization of the
  spectral problem) \emph{Int. J. Mod. Phys. A} \textbf{4} 2897--2952

\bibitem{GKO94}
Gonz\'{a}lez-L\'{o}pez A, Kamran N and Olver P~J 1994 Quasi-exact solvability
  \emph{Contemp. Math.} \textbf{160} 113--140

\bibitem{Cr55}
Crum M 1955 Associated {S}turm--{L}iouville equations \emph{Quart. J. Math.
  Oxford} \textbf{6} 121--127

\bibitem{Kr57}
Kre\v\i{}n M 1957 \emph{Dokl. Akad. Nauk SSSR} \textbf{113} 970

\bibitem{GKO93}
Gonz\'{a}lez-L\'{o}pez A, Kamran N and Olver P~J 1993 Normalizability of
  one-dimensional quasi-exactly solvable {S}chr{\"o}dinger operators
  \emph{Commun. Math. Phys.} \textbf{153} 117--146

\bibitem{Tu94}
Turbiner A 1994 {L}ie algebras, cohomologies, and new findings in quantum
  mechanics \emph{Contemp. Math.} \textbf{160} 263--310

\bibitem{Tu88}
Turbiner A 1988 Quasi-exactly solvable problems and {$\fsl(2)$} algebra
  \emph{Commun. Math. Phys.} \textbf{118} 467--474

\bibitem{Tu92}
Turbiner A 1992 {L}ie algebras and polynomials in one variable \emph{J. Phys.
  A: Math. Gen.} \textbf{25} L1087--L1093

\bibitem{CKS95}
Cooper F, Khare A and Sukhatme U 1995 Supersymmetry and quantum mechanics
  \emph{Phys. Rep.} \textbf{251} 267--385 \eprint{hep-th/9405029}

\bibitem{GR00}
Gradshteyn I and Ryzhik I 2000 \emph{{T}able of {I}ntegrals, {S}eries, and
  {P}roducts} (San Diego: Academic Press) sixth edition

\bibitem{DF98}
Dunne G and Feinberg J 1998 Self-isospectral potentials and supersymmetric
  quantum mechanics \emph{Phys. Rev. D} \textbf{57} 1271--1276
  \eprint{hep-th/9706012}

\bibitem{BD96}
Bender C~M and Dunne G~V 1996 Quasi-exactly solvable systems and orthogonal
  polynomials \emph{J. Math. Phys.} \textbf{37} 6--11 \eprint{hep-th/95111389}

\bibitem{FGR96}
Finkel F, Gonz{\'a}lez-L{\'o}pez A and Rodr{\'\i}guez M~{\'A} 1996
  Quasi-exactly solvable potentials on the line and orthogonal polynomials
  \emph{J. Math. Phys.} \textbf{37} 3954--3972 \eprint{hep-th/9603103}

\bibitem{Ch78}
Chihara T 1978 \emph{{A}n {I}ntroduction to {O}rthogonal {P}olynomials} (New
  York: Gordon and Breach)

\bibitem{TFC03}
Tsutsui I, F{\"u}l{\"o}p T and Cheon T 2003 Connection conditions and the
  spectral family under singular potentials \emph{J. Phys. A: Math. Gen.}
  \textbf{36} 275--287 \eprint{quant-ph/0209110}

\bibitem{Ki03}
Kikuchi H 2003 The fermion determinant, its modulus and phase \emph{Phys. Lett.
  B} \textbf{562} 299--306 \eprint{hep-th/0210003}

\bibitem{Sm03}
Smilga A 2004 Weak supersymmetry \emph{Phys. Lett. B} \textbf{585} 173--179
  \eprint{hep-th/0311023}

\end{thebibliography}
\bibliographystyle{jpa}

\end{document}